\documentclass[fleqn,usenatbib]{mnras}

\usepackage{newtxtext,newtxmath}

\usepackage[T1]{fontenc}

\DeclareRobustCommand{\VAN}[3]{#2}
\let\VANthebibliography\thebibliography
\def\thebibliography{\DeclareRobustCommand{\VAN}[3]{##3}\VANthebibliography}

\usepackage{graphicx}	
\usepackage{amsmath}	
\usepackage{booktabs}
\usepackage{ulem}
\usepackage{orcidlink}

\newcommand{\lhgpc}{{h^{-1}\rm Gpc}}
\newcommand{\lhmpc}{{h^{-1}\rm Mpc}}
\newcommand{\lhkpc}{{h^{-1}\rm kpc}}
\newcommand{\msun}{{h^{-1}\rm M_{\odot}}}

\title[]{Reproducing Abell 2744 with the HyperMillennium Simulation}

\author[Wang \& Li et al.]{
Qiao Wang\orcidlink{0000-0003-2153-7758},$^{1,2}$\thanks{These three authors contributed equally to this work.}\thanks{E-mail: qwang@nao.cas.cn (QW); mingli@nao.cas.cn (ML); lgao@bnu.edu.cn (LG)}
Ming Li\orcidlink{0000-0002-1318-4828},$^{1,2}$\footnotemark[1]\footnotemark[2]
Liang Gao\orcidlink{0009-0006-3885-9728},$^{3,4}$\footnotemark[2]
Qi Guo,$^{3}$\footnotemark[1]
Raul E. Angulo\orcidlink{0000-0003-2953-3970},$^{5,6}$
Sangjun Cha\orcidlink{0000-0001-7148-6915},$^{7}$
\newauthor
Shaun Cole\orcidlink{0000-0002-5954-7903},$^{8}$
Carlos S. Frenk,$^{8}$
Kim HyeongHan\orcidlink{0000-0002-2550-5545},$^{7,9}$
Ran Li\orcidlink{0000-0003-3899-0612}, $^{3}$
Wenxiang Pei\orcidlink{0009-0000-0690-5562},$^{10}$
Huanyuan Shan\orcidlink{0000-0001-8534-837X},$^{11,12,2}$
\newauthor
Jie Wang\orcidlink{0000-0002-9937-2351}, $^{1,2}$
and Simon~D.~M. White\orcidlink{0000-0002-1061-6154}$^{13}$
\\
$^{1}$ National Astronomical Observatories, Chinese Academy of Sciences, 20A Datun Road, Chaoyang District, Beijing 100101, China \\
$^{2}$ School of Astronomy and Space Science, University of Chinese Academy of Sciences, Beijing 100048, China \\
$^{3}$ Institute for Frontiers in Astronomy and Astrophysics, Beijing Normal University, Beijing 102206, China \\
$^{4}$ School of Physics and Microelectronics, Zhengzhou University, Zhengzhou 450001, China \\
$^{5}$ Donostia International Physics Center, Manuel Lardizabal Ibilbidea, 4, 20018 Donostia, Gipuzkoa, Spain \\
$^{6}$ IKERBASQUE, Basque Foundation for Science, 48013, Bilbao, Spain \\
$^{7}$ Department of Astronomy, Yonsei University, 50 Yonsei-ro, Seoul 03722, Korea \\
$^{8}$ Institute for Computational Cosmology, Department of Physics, Durham University, South Road, Durham DH1 3LE, UK \\
$^{9}$ Department of Physics, Duke University, Durham, NC 27708, USA \\
$^{10}$ Shanghai Key Lab for Astrophysics, Shanghai Normal University, Shanghai 200234, China \\
$^{11}$ Shanghai Astronomical Observatory, Chinese Academy of Sciences, Shanghai 200030, China \\
$^{12}$ State Key Laboratory of Radio Astronomy and Technology, 20A Datun Road, Chaoyang District, Beijing 100101, China \\
$^{13}$ Max Planck Institute for Astrophysics, Karl-Schwarzschild-Strasse 1, 85740 Garching, Germany
}

\date{Accepted XXX. Received YYY; in original form ZZZ}

\pubyear{2025}

\begin{document}
\label{firstpage}
\pagerange{\pageref{firstpage}--\pageref{lastpage}}
\maketitle


\begin{abstract}
We present the Hyper Millennium (HM) simulation, an extremely large cosmological simulation designed to support next-generation galaxy surveys. The simulation follows 4.2 trillion dark matter particles in a comoving box of $2.5\ \lhgpc$, with a mass resolution of $3.2 \times 10^8\, \msun$ and a force resolution of $3.0\ \lhkpc$. Its combination of scale and resolution is ideal for studying large-scale structures and rare cosmic objects. In this first paper of the HM project, we explore whether the massive galaxy cluster Abell~2744 (A2744) can be reproduced in detail in the simulation. Pixel-based statistics of galaxy number density $N_{\rm gal}$, luminosity density $L_{\rm gal}$, and projected mass density $\kappa$ show excellent agreement between A2744 and its analogues down to $\sim 50$~kpc, once field-selection biases toward high galaxy surface density are accounted for. This concordance, achieved in one of the most extreme known galaxy environments, is a validation of the underlying $\Lambda{\rm CDM}$ model in the extreme regime of A2744. It also showcases the robustness and accuracy of the HM simulation, which, when coupled with a sophisticated semi-analytic galaxy formation model, is capable of producing galaxy and mass catalogues of comparable quality out to high redshift across its full comoving volume of 50.4 ${\rm Gpc^3}$.
\end{abstract}



\section{Introduction}
Over the past two decades, significant advancements in both theoretical and observational astronomy have culminated in the establishment of the ``standard'' cosmological model, the $\Lambda{\rm CDM}$ model (cold dark matter with a cosmological constant). This model has been remarkably successful in reproducing a wide range of observations, from the temperature fluctuations in the cosmic microwave background to the large-scale distribution of galaxies in the present-day Universe. In particular, large $N$-body simulations of structure formation have played a pivotal role, with notable examples including the Millennium \citep{MS_06}, Millennium-II \citep{MSII_09}, Millennium-XXL \citep{Angulo12}, Aquarius \citep{Aquarius_08}, Phoenix \citep{Phoenix_12}, and more recent efforts such as the BACCO \citep{Bacco_20} and simulations by \citet{WangJ_20}. These simulations have provided critical insights into the evolution of cosmic structures across a range of scales. 

Despite these achievements, fundamental questions about the Universe remain unresolved, including the nature of dark matter and dark energy. Current and upcoming wide-field galaxy surveys, both ground- and space-based, are poised to address these questions with unprecedented precision. Projects such as the \textit{Dark Energy Survey} \citep[DES,][]{DES_16}, the \textit{Dark Energy Spectroscopic Instrument} \citep[DESI,][]{DESI_22}, \textit{Euclid} \citep{Euclid_11}, the \textit{Chinese Space Station Telescope} \citep[CSST,][]{CSST_11}, the \textit{Large Synoptic Survey Telescope} \citep[LSST,][]{LSST_19}, and the \textit{Nancy Grace Roman Space Telescope} \citep[\textit{Roman,}][]{Spergel_13} are expected to measure cosmological parameters to percent-level precision, providing a transformative boost to our understanding of these fundamental issues.

Modern hydrodynamical galaxy formation simulations, such as EAGLE \citep{Eagle_15a,Eagle_15b}, Illustris \citep{Illustris_14b,Illustris_14a}, IllustrisTNG \citep{IllustrisTNG_17,IllustrisTNG_18}, and more recent projects like MillenniumTNG \citep{MTNG_23a,MTNG_23b} and FLAMINGO \citep{Flamingo_23}, have significantly advanced our ability to reproduce small-scale observations by incorporating complex baryonic physics. However, large pure dark matter simulations remain indispensable for interpreting the results of forthcoming galaxy surveys. These simulations are crucial for understanding the complex physical processes underlying various cosmological probes, such as baryon acoustic oscillations (BAO), redshift-space distortions (RSD), and weak lensing. They also play a key role in quantifying selection effects, systematic biases, and statistical uncertainties in large-scale surveys. Given the vast sky coverage of these surveys, the required simulation volumes must be extremely large—on the order of $(2\ \lhgpc)^3$ or more—to accurately measure the power spectrum on scales of $\sim100\ \mathrm{Mpc}$. Furthermore, the ability to resolve faint objects, such as emission-line galaxies, necessitates a mass resolution capable of resolving dark matter halos down to $\sim10\%$ of the Milky Way's mass, corresponding to a particle mass of $\sim10^8\msun$. State-of-the-art simulations, such as Uchuu \citep{Uchuu_2021} and the Euclid Flagship II \citep{Euclid_FL2}, have been designed to meet these stringent requirements.

In this paper, we present the HyperMillennium (HM) simulation, an exceptionally large cosmological $N$-body simulation designed to address the needs of new generation of galaxy surveys. The HM simulation features twice the particle count of the Uchuu simulation, while maintaining comparable mass resolution. Compared to the Euclid Flagship II simulation, HM achieves superior resolution and includes 100 output snapshots, enabling detailed tracking of dark matter halo growth histories. This capability is critical for constructing reliable mock galaxy catalogues tailored to modern surveys. With an unprecedented volume of $(2.5\ \lhgpc)^3$, the HM simulation is also ideally suited for studying rare cosmic objects, such as massive galaxy clusters. For instance, Abell 2744, a complex system with puzzling substructures, has posed challenges to the $\Lambda{\rm CDM}$ paradigm. The HM simulation, with its combination of large volume and high resolution, provides a unique opportunity to investigate whether such systems are consistent with theoretical expectations. These topics form the main focus of this paper.

Our paper is organised as follows: In Section 2, we present the details of the HyperMillennium (HM) simulation, including the computational code used to perform the run and the semi-analytical galaxy formation model employed to populate galaxies. Section 3 provides an overview of the basic statistical properties derived from the simulation. In Section 4, we describe the methodology for generating JWST mock observations of Abell 2744. The main results of our study are discussed in Section 5, and the paper concludes with a summary of our findings in Section 6.

\section{Numerical methods}
\subsection{The Hyper-Millennium simulation}

\begin{figure*}
\centering
\includegraphics[width=1.0\textwidth]{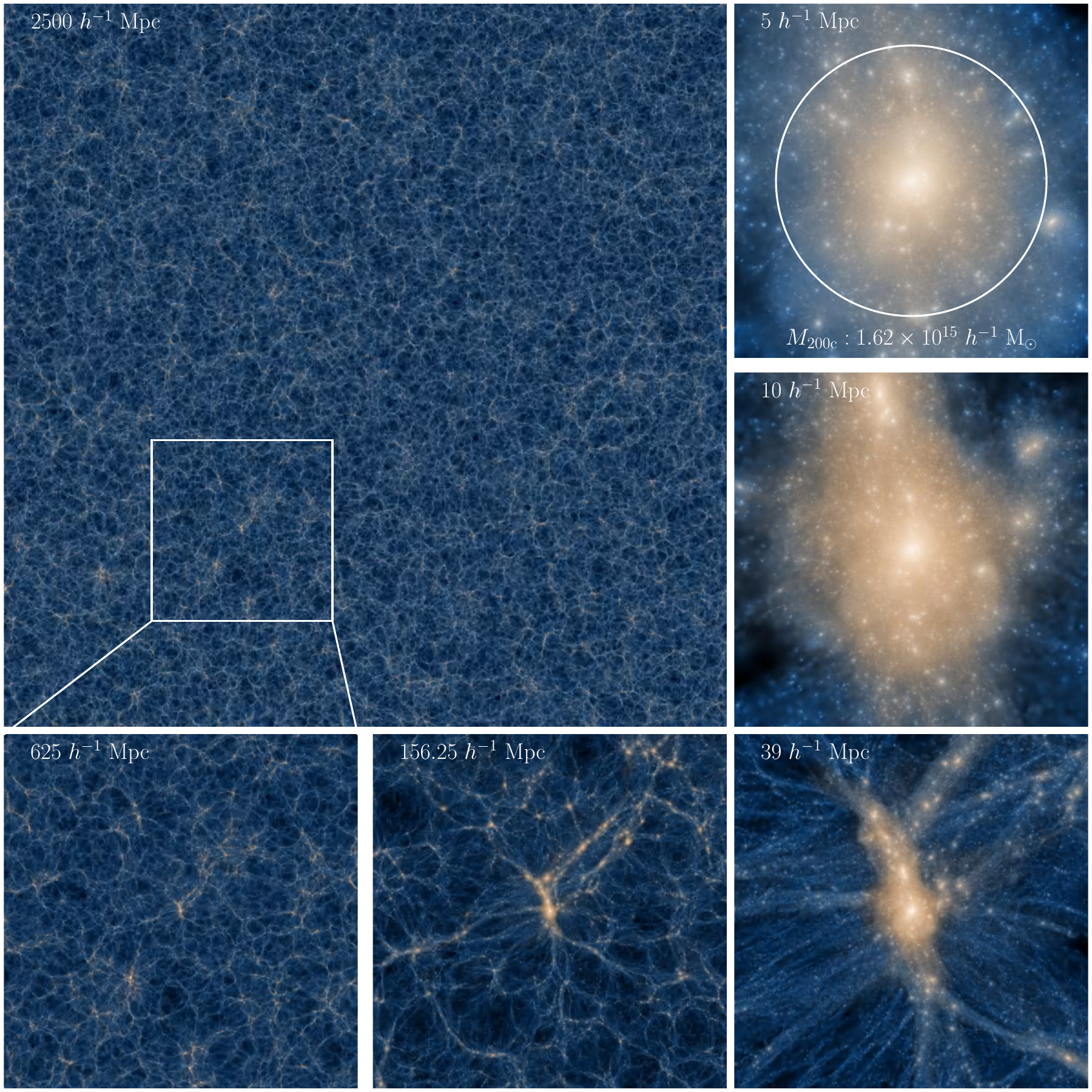}
\caption{ Dark matter density slice from the HM simulation at the present epoch ($z = 0.0$). The upper-left panel displays the full simulation box with a side length of 2.5~$h^{-1}$~Gpc. The subsequent panels, arranged counterclockwise, progressively zoom in on a massive dark halo across five levels, from the large-scale structure (625~$h^{-1}$~Mpc) down to sub-halos within the inner regions of dark halos (5~$h^{-1}$~Mpc).
}\label{fig:snapshot}
\end{figure*}

\begin{figure}
\centering
\includegraphics[width=0.47\textwidth]{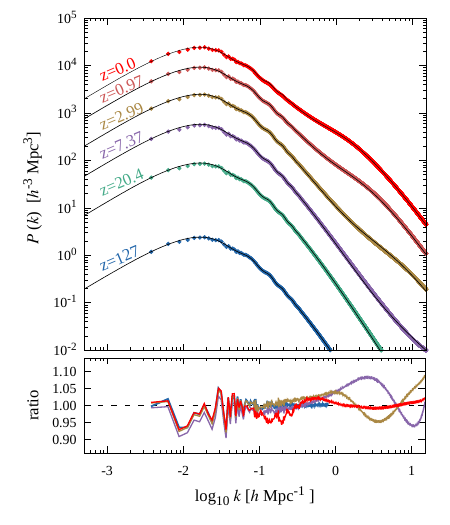}
\caption{ The power spectrum evolution from the initial redshift ($z = 127$) to the present day ($z = 0$). The solid curves represent the reference predictions from CAMB with HMCode \citep{2016MNRAS.459.1468M}. The lower panel shows the ratio of the simulation measurements to the predictions, highlighting any deviations.
}\label{fig:power}
\end{figure}

The Hyper-Millennium (HM) simulation evolves approximately 4.2 trillion ($16128^3$) dark matter particles within a cubic volume of side length $2.5\ \lhgpc$. The number of particles in the HM simulation is twice that of the state-of-the-art Uchuu simulation \citep[]{Uchuu_2021} and slightly exceeds that of the latest Euclid Flagship II simulation \citep[]{Euclid_FL2}. The mass resolution of the HM simulation is $3.2 \times 10^8\ \msun$, comparable to that of Uchuu but significantly higher than that of our previous large-volume Millennium-XXL (MXXL) simulation ($6.2 \times 10^9\ \msun$) and the Euclid Flagship II simulation ($1.0 \times 10^9\ \msun$). 
The force resolution is set to $\epsilon = 3.0\ \lhkpc$. For the softening scheme, we utilise a quadratic function to set up the softening potential, instead of using a Plummer shape. So that the gravity $g(r)$ at the softening scale reads
\begin{equation}\notag
     g(r) \sim \left\{ 
     \begin{matrix}
        - {G r }/{ h^3 } &(r \leq h)\\
        - {G }/{ r^2 }  & ( r > h)\\
     \end{matrix}
     \right.,
\end{equation}
where the boundary radius is $h = 1.5 \epsilon$.
The simulation adopts the cosmological parameters from the Planck mission: $H_0 = 67.66\ {\rm km\ s^{-1}\ Mpc^{-1}}, \Omega_{\Lambda} = 0.6889,\ \Omega_{\rm m} = \Omega_{\rm dm} + \Omega_{\rm b} = 0.3111, \Omega_{\rm b} = 0.049, \sigma_8 = 0.8102$, and $n_{\rm s} = 0.9665$.

The initial condition (IC) of the simulation was generated using a built-in module based on the Zel'dovich approximation, assuming that the total matter distribution follows the linear power spectrum computed by the CAMB code. 
Recent studies have shown that the use of late-start time ICs based on higher-order Lagrangian perturbation theory (LPT) would help suppress potential particle noise \citep[]{Michaux_Etal_2021,Angulo_Hahn_2022}. Due to memory constraints, we were unable to employ 2nd-order LPT. To ensure sufficient accuracy for the Zel'dovich approximation in capturing the non-linear evolution of structures, the simulation starts from an early redshift of $z=127$.

The HM simulation outputs 100 snapshots from $z=20.4$ to $z=0$, following a timing scheme similar to that of the Millennium-II simulation \citep[see Eq. 2 in][]{MSII_09}. The total disk storage required for the particle data is approximately $12$ PB. Due to memory limitations, distributed post-processing was performed after the simulation run, including the identification of groups and subhalos. Dark matter halos containing at least 20 particles were identified using a standard friends-of-friends (FOF) algorithm with a linking length of 0.2 times the mean particle separation. At $z=0$, the HM simulation resolves approximately $8 \times 10^9$ FOF groups. Subhalos within FOF groups were further identified using the SUBFIND algorithm. The subhalo catalogue requires an additional $868$ TB of storage.

Using the subhalo catalogue, merger trees were constructed to track subhalo evolution between adjacent snapshots, following the methods described in \citet{Angulo12} and \citet{Angulo14}. The merger trees, along with auxiliary files for implementing semi-analytic galaxy formation models, occupy approximately $150$ TB of storage, or 170 TB when including progenitor/descendant link files.

\subsection{The Code}
The HM simulation is performed with the latest version of the \texttt{PhotoNs-3.7} code, based on the original version described in \citet{Wang_2018}. The code adopts a hybrid particle mesh (PM) and Fast Multipole Method (FMM) scheme to compute gravity \citep{Wang_2021, 2021MNRAS.506.2871S}. Compared to the popular PM-Tree method, PM-FMM has the attractive feature of a time complexity of $O(N)$, making it ideal for performing extremely large simulations, such as the HM run, on massively parallel supercomputers. 

Given the dominance of heterogeneous architectures in current high-performance computing (HPC), the \texttt{PhotoNs} code is specifically designed to leverage the power of heterogeneous HPC machines. In practice, the most computationally expensive operator, particle-to-particle (P2P), which accounts for more than $90$\% of the total calculations, is offloaded to GPU cores (Graphics Processing Units; see \citet{Wang_2021_gpu} for technical details). The \texttt{PhotoNs} code was previously used to perform the Ultramarine simulation, a 2-trillion-particle run designed to study structure formation in the high-redshift universe \citep{2022MNRAS.517.6004W}. For the more challenging HM simulation, we rewrote the computing kernels, fine-tuned their instructions, and adjusted the parameters in registers and memory to adapt to the specific GPGPU (General Purpose Graphics Processing Units) architecture of the ORISE supercomputer. Meanwhile, the domain decomposition and MPI communication are reconstructed for the HM simulation.

Due to the memory constraints of running 4 trillion particles, gravity was evaluated using 32-bit single-precision floating-point numbers in the HM simulation. Single precision is sufficient for most calculations except for particle positions. IEEE 32-bit floating-point precision has a relative precision of up to $2^{23} \sim 10^{-7}$, whereas the spatial resolution of the HM run requires about $3/2500000 \sim 1.2 \times 10^{-6}$. This discrepancy is not small enough to accurately record positions. Test runs demonstrated that using single floating-point precision artificially reduces the abundance of low-mass halos, particularly in the diagonal direction of the simulation box. To address this issue, we recorded particle positions using 32-bit integer numbers ($\sim 10^{-10}$) instead of IEEE floating-point numbers in memory and on a hard disk. When position information was needed, the integers were transformed into sufficient precision, which was then used for computing spatial separations, mesh construction, and other tasks. After the calculations were complete, an incremental integer for the particle drift was accumulated to record particle positions. To validate this mixed precision and quantisation, we ran a small simulation with $768^3$ particles and compared the results with those obtained using the \texttt{GADGET} code under the same computational conditions. The comparison showed consistent results for the positions of dark matter halos and subhalos.

Before performing the main simulation, we conducted a series of $2592^3$ simulations using both the \texttt{GADGET-4} and \texttt{PhotoNs} codes. These tests showed very good agreement over a broad set of statistics in numerical consistency, including the power spectrum, halo/subhalo mass functions, and the internal structures of dark matter halos. The HM simulation required approximately $420$ wall-clock hours and was completed in over $12,000$ time steps.

\begin{figure}
\centering
\includegraphics[width=0.47\textwidth]{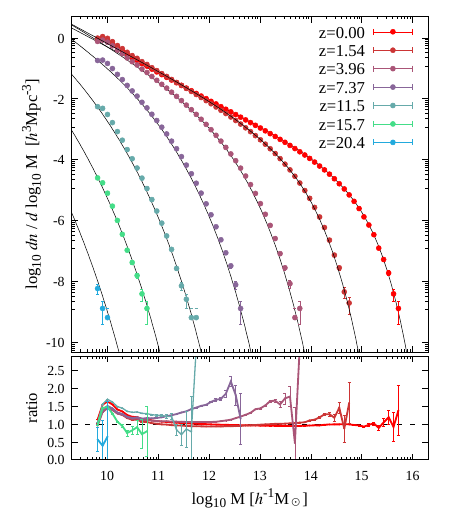}
\caption{ The FOF halo mass functions measured at seven redshifts, from $z = 20.4$ to $z = 0$. The minimum halo, containing 20 particles, corresponds to a mass of $\sim 6.5 \times 10^{9}~h^{-1}\mathrm{M}_\odot$. The solid curves represent the predictions from the fitting model of \citet{2007MNRAS.374....2R}. The bottom panel shows the ratio as a function of halo mass, comparing the simulation results with the model predictions.
}\label{fig:halo}
\end{figure}

\subsection{The semi-analytic model of galaxy formation}
\label{sec:sam_model}
The semi-analytic galaxy formation model (SAM) is an efficient and well-established approach for populating galaxies in dark matter only simulations. It has been demonstrated to successfully reproduce various observed galaxy properties, both in the local Universe and at high redshifts \citep[e.g.,][]{Kauffmann_99,Cole_2000,Guo_11,Lacey_16}. 

The SAM framework tracks the evolution of galaxies within the structure provided by the dark matter halo merging trees. It includes parameterised recipes for key galaxy formation processes, such as gas cooling, star formation, reionisation heating, supernova feedback, mergers, black hole growth, metal enrichment, and feedback from active galactic nuclei (AGN). In this work, we employ an updated version of the SAM code \texttt{L-GALAXIES}, incorporating galaxy formation prescriptions based on \citet{2015MNRAS.451.2663H} and \citet{Pei_etal_24}, which has been previously calibrated to simultaneously reproduce a broad range of key observational constraints, including the stellar mass function (SMF), galaxy colour distributions, and the black hole mass function (BHMF). We do not re-tune the model prescriptions specifically for the HM simulation. This ensures that our results are not overfitted to the specific object (A2744) under study.

By combining galaxy formation histories derived from \texttt{L-GALAXIES} with a stellar synthesis model, we compute the full spectral energy distribution (SED) for each individual galaxy. In particular, we further generate predicted galaxy magnitudes across multiple filters used by the James Webb Space Telescope (JWST). After populating selected dark matter halos from our HM simulation, as described in Section~\ref{sec:HM_sim_a2744}, we create realistic visualisations that align closely with observational data, using these predicted galaxy catalogues.

\section{Basic statistical results of the simulation }

In Figure~\ref{fig:snapshot}, we present a slice of the dark matter density field. The upper-left panel displays the full scale of the simulation box, while the remaining panels progressively zoom into smaller scales in an anticlockwise direction. The top-right panel highlights a very massive, rich cluster halo with a halo mass of $M_{\rm 200c} = 1.62 \times 10^{15}\ \msun$. This visualisation illustrates the vast dynamical range resolved by the HM simulation, with the density field appearing nearly uniform on ${\rm Gpc}$ scales and highly clumpy on ${\rm sub-Mpc}$ scales.

Figure~\ref{fig:power} shows the power spectrum of the dark matter density field at seven redshifts: $z = 127, 20.4, 7.37, 2.99, 0.97, 0$. The power spectrum is computed on the fly using a regular $16128^3$ Cloud-in-Cell (CIC) mesh. For reference, we overplot theoretical predictions from the HMCode formula \citep{2016MNRAS.459.1468M}. To facilitate comparison, we also show the ratio of the simulation measurements to the predictions. Remarkably, the simulation results agree closely with the theoretical predictions across all scales and redshifts shown here, with a maximum deviation of only about $5$ per cent.

\begin{figure*}
\centering
\includegraphics[width=1.0\textwidth]{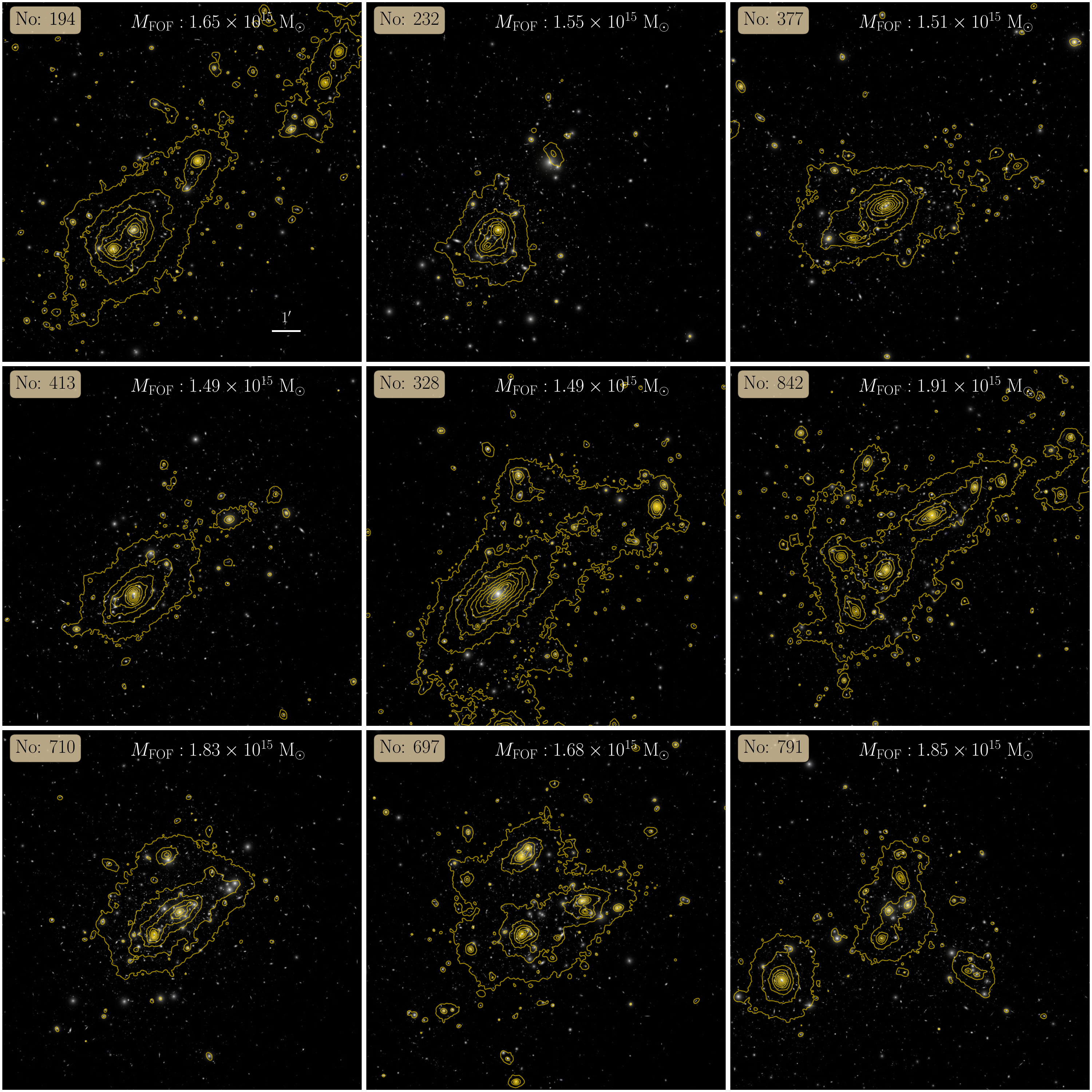}
\caption{A2744 analogue clusters selected from the HM simulation as they would appear at a redshift of $z = 0.308$. A composite-colour image with a field coverage of $12'.5 \times 12'.5$ for each cluster is shown, using the F444W filter for red, F277W for green, and F115W for blue. The yellow contours represent the projected mass distribution, smoothed here with a Gaussian kernel of $\sigma=2^{\prime\prime}$, mimicking the lensing convergence.}
\label{fig:mock_a2744_combo}
\end{figure*}

Due to the large volume and high mass resolution, the HM simulation resolves approximately $8 \times 10^{9}$ (8,358,635,412) FOF groups with a minimum of 20 particles. Among these groups, more than $480,184$ have masses greater than $10^{14}\ \msun$, and approximately $3,740$ are more massive than $10^{15}\ \msun$. The most massive group has a total mass of $M_{\rm FOF} = 4.9 \times 10^{15}\ \msun$ and a spherical overdensity mass of $M_{\rm 200c} = 3.6 \times 10^{15}\ \msun$. 

In Figure~\ref{fig:halo}, we show the differential FOF halo mass functions measured at seven redshifts, ranging from $z=20.4$ to $z=0$. The minimum halo mass corresponds to 20 particles, or approximately $6.5 \times 10^{9}\ \msun$. At $z=20.4$, the HM simulation resolves $12$ halos more massive than $10^9\ \msun$, which are as rare at that epoch as the most massive objects in the Universe are today. At $z=15.7$, corresponding to the redshift of the earliest galaxies that the JWST may already observe, the simulation still resolves a few hundred haloes with abundances comparable to that of the Coma cluster at the present day.

The solid lines in Figure~\ref{fig:halo} represent the fitting model from \citet[]{2007MNRAS.374....2R}, while the bottom panel shows the ratio between the model predictions and the simulation measurements. Overall, the model agrees remarkably well with the simulation across most mass scales at $z=0$. However, the agreement worsens at the high-mass end for redshifts $z=1.54$, $3.96$, and $7.37$, with deviations reaching up to a factor of $1.5$ in regions where the simulation has excellent statistics and a large number of particles. At $z=20.4$ and $z=15.7$, the measurements roughly align with the model predictions, albeit with considerable noise. It is worth noting that the small bumps in the mass functions around $10^{11}\ \msun$ are purely numerical effects caused by the discreteness of low particle counts.

\section{Abell 2744 analogues in the HM simulation}
\label{sec:HM_sim_a2744}
The A2744 galaxy cluster is a peculiar and massive object apparently undergoing multiple merging events involving several individual components with masses exceeding $10^{14}\ \msun$. It has been extensively studied through both ground- and space-based observational programmes. The hierarchical clustering and merging processes that lead to the observed configuration of galaxies, gas, and dark matter are highly complex. Gravitational lensing models provide an unbiased way to image the total mass distribution, encompassing both luminous and dark components. 

By combining weak lensing (WL) and strong lensing (SL) techniques, the mass distribution of A2744 has been recovered in exceptional detail from the inner core of the cluster to its virial radius, extending even to the surrounding large-scale structure. We leverage the exceptional statistics and high mass resolution of the HM simulation to search for close analogues of A2744. By comparing simulated analogues to the observational data, we aim to determine whether the detailed, multi-scale relation between structures in the mass and galaxy distributions, which is evident in the A2744 lensing maps, can be reproduced within the $\Lambda$CDM framework.

\subsection{Selecting A2744 analogues}
\label{sec:proc_analysis}
A weak-lensing analysis of A2744 based on Subaru/Suprime-Cam imaging \citep[]{2016ApJ...817...24M} estimated a virial mass of $2.06 \pm 0.42 \times 10^{15}\ {\rm M_{\odot}}$. Several recent studies have used the JWST/NIRCam imaging to map the total mass distribution of the cluster. For example, \citet[their Figure 8]{Bergamini_Etal_23} present the projected mass of A2744, showing it to exceed $10^{15}\ {\rm M_{\odot}}$ in the region enclosing the five brightest galaxies. Likewise, \citet[Figure 6]{Cha_Etal_2024} estimate a total projected mass of $\sim1.19\times 10^{15}\ {\rm M_{\odot}}$ within a $1\ {\rm Mpc}$ aperture centred on their lensing reconstruction field.

Motivated by these findings, we first restrict our halo sample to systems with $M_{\rm FOF} > 10^{15}\ \msun$, yielding 851 FOF groups in snapshot  082 at redshift $z=0.300$, close to that of A2744 ($z_{\rm cluster}=0.308$). Because A2744 hosts multiple bright galaxies, we further refine the sample by comparing our simulated halo configurations to the observed positions of the five most prominent structures in A2744: these surround the northern brightest cluster galaxy (BCG-N), the southern brightest cluster galaxy (BCG-S), and three additional massive galaxies, G1, G2, and G3. For each simulated halo in the sample, we implement an iterative projection algorithm that explores the relative positions of the five most massive substructures from multiple viewing angles. To quantify the similarity between the positions of the simulated clumps and those observed in A2744, we make use of \texttt{Procrustes} analysis \citep{Gower_04,Dryden_16}\footnote{\url{https://en.wikipedia.org/wiki/Procrustes_analysis}}. 

\texttt{Procrustes} is a well-established geometric morphometry technique that finds the optimal transformation between two sets of corresponding points. In our application, we rank the five A2744 bright galaxies in descending order by their lensing reconstructed masses \citep[][see, e.g. Table\ 2]{Cha_Etal_2024}. Their observed sky positions are denoted as $\mathbf{P}=\left\{\mathbf{p}_1, \mathbf{p}_2, \ldots, \mathbf{p}_5\right\}$, where each $\mathbf{p}_i$ represents the two-dimensional sky coordinates of galaxy $i$. Similarly, for each candidate halo in the sample, we apply a random three-dimensional (3D) rotation and project its five most massive substructures onto a two-dimensional (2D) plane, yielding 2D positions $\mathbf{Q}=\left\{\mathbf{q}_1, \mathbf{q}_2, \ldots, \mathbf{q}_5\right\}$, with $\mathbf{q}_i$ ordered by their masses.

We then solve the optimal combination of 2D rotation $\mathbf{R}$, translation $\mathbf{t}$, and uniform scaling $\mathbf{s}$ that minimises the sum of squared distances between the two corresponding sets of points,
\begin{equation}
    \label{eq:proc_distance}
    D_{\rm proc}^2 = \sum_{i=1}^{5} \|\mathbf{p}_i - (\mathbf{s}\mathbf{R}\mathbf{q}_i + \mathbf{t})\|^2 \ .
\end{equation}
$\mathbf{R}$ is a $2\times2$ rotation matrix. The translation vector $\mathbf{t}$ is introduced to align the centroids of the two point sets. As a result of including the scaling factor $\mathbf{s}$, only the shape of the sky configuration of the five clumps is required to match that in A2744, not the overall size; 
we do not, however, scale the simulated halos when generating virtual images and galaxy number and luminosity maps in the following sections.

For each candidate halo, this process is repeated multiple times over random orientations. The \texttt{Procrustes} distance $D_{\rm proc}$ is then minimised to find the best match of projection that yields the minimum value $D_{\rm proc, min}$. We then rank all candidates in ascending order of their $D_{\rm proc, min}$ and select those whose projected substructure configurations most closely resemble A2744 (detailed descriptions can be found in Appendix \ref{app:Procrustes}). 

In this work, we present two A2744 analogue samples. \textbf{Group I} consists of candidates determined purely by geometrical similarity (Appendix \ref{app:PGrp1}); while \textbf{Group II} consists of candidates for which the similarity of the structure mass ratios was also used as a selection criterion (Appendix \ref{app:PGrp2}). 
The candidate IDs, FOF masses, and other properties are summarised in Table~\ref {tab:DM_analogues}, along with relevant comparison properties of A2744. It is notable that for the clusters in {\bf Group II}, the $M_{\rm 200c}$ values are typically substantially smaller both than those of {\bf Group I} clusters and than their own FOF halo masses. This is because reproducing both the geometric pattern and the mass ratios of the A2744 substructures has resulted in a preference for systems in which multiple massive substructures (that might be undergoing merging events) are projected close to each other along the line of sight.

\subsection{Constructing mock Observational data}
\subsubsection{The projected mass distribution}
\label{sec:kappa_maps}
To make a direct comparison with the lensing reconstruction of A2744, projected mass distribution maps are required for our candidate halos. In this subsection, we outline the procedure we follow. For each candidate, we centre a rectangular box with a side of $3\ \lhmpc$ and a depth of $\pm20\ \lhmpc$ on the halo's potential minimum and orient it along the direction defined in the previous section. All simulation particles within this box are then projected onto a plane perpendicular to the presumed orientation. When placed at the redshift of $z=0.308$, the resulting projected mass maps cover a region of $12'.5 \times 12'.5$, which we sample with $750 \times 750$ pixels to achieve a resolution of 1 arcsec, to be the same as the resolution of the mass map of \citet{Cha_Etal_2024}. 

The projected mass distributions are then converted to convergence maps via the relation,
\begin{equation}
    \kappa = \Sigma/\Sigma_{\rm c},
\end{equation}
where $\Sigma$ is the surface mass density, and $\Sigma_{\rm c}$ is the critical surface mass density, defined as:
\begin{equation}
    \Sigma_{\rm c} = \frac{c^2 D_{\rm s}}{4\pi G D_{\rm d} D_{\rm ds}},
\end{equation}
where $c$ is the speed of light, $D_{\rm d}$ is the angular diameter distance to the lens, $D_{\rm s}$ is the angular diameter distance to the source, and $D_{\rm ds}$ is the angular diameter distance between the lens and the source.  To facilitate comparison with the reconstructed observational data, we present the $\kappa$ field estimated using a critical surface density of $1.722 \times 10^9\ {\rm M}_{\odot}\ {\rm kpc}^{-2}$ (corresponding to the value when setting $D_{\rm ds} / D_{\rm d}=1$); this differs slightly from the $1.777 \times 10^9\ {\rm M}_{\odot}\ {\rm kpc}^{-2}$ adopted by \citet{Cha_Etal_2024} because of differences in the assumed cosmological parameters.

\begin{table*}
    \caption{The properties of A2744 analogues in HM simulation.}
    \begin{tabular}{l c c c c c c l c c c c c}
    \hline
    \multicolumn{5}{c}{\textbf{Group I}}  && \multicolumn{6}{c}{\textbf{Group II}}  \\
    \cline{1-6}   \cline{8-13} \\

    No. & $M_{\rm FOF}$              & $M_{\rm 200c}$             & $M_{\rm 2D}$               & ${N_{\rm gal}}$ & $\mathcal{R}$ && 
    No. & $M_{\rm FOF}$              & $M_{\rm 200c}$             & $M_{\rm 2D}$               & ${N_{\rm gal}}$ & $\mathcal{R}$ \\
        & $[10^{14}\ \rm M_{\odot}]$ & $[10^{14}\ \rm M_{\odot}]$ & $[10^{14}\ \rm M_{\odot}]$ &   &  &&     
        & $[10^{14}\ \rm M_{\odot}]$ & $[10^{14}\ \rm M_{\odot}]$ & $[10^{14}\ \rm M_{\odot}]$ &   & \\
    (1) & (2) & (3) & (4)$^{\rm a}$  & (5)$^{\rm b}$ & (6)$^{\rm c}$ && (7) & (8) & (9) & (10)$^{\rm a}$ & (11)$^{\rm b}$  & (12)$^{\rm c}$ \\
    \cline{1-6}   \cline{8-13} \\

    194 & 16.5  & 15.8 & 11.6 & 931 (564) & 2.6 &&    842 & 19.1  & 4.5 &  14.5  & 1082 (754)  & 1.2 \\
    232 & 15.5  & 15.3 & 11.6 & 689 (509) & 2.5 &&    710 & 18.3  & 9.0 &  16.1  & 1021 (807)  & 1.8 \\
    377 & 15.1  & 13.7 & 10.8 & 833 (534) & 2.9 &&    697 & 16.8  & 9.3 &  12.7  & 906  (636)  & 1.5 \\
    413 & 14.9  & 13.2 & 10.2 & 779 (484) & 3.0 &&    791 & 18.5  & 7.0 &  10.8  & 1022 (574)  & 1.2 \\
    328 & 14.9  & 14.1 & 10.5 & 770 (486) & 3.1 &&  & & & & \\
    \hline
    ${\textbf{A2744}}^{\dagger}$ & --- & \textbf{20.6} $\pm$ \textbf{4.2} & \textbf{10.9} & \textbf{768}$^{\ddagger}$ \textbf{(504)}  &&  & & & &   \\

    \hline

    \end{tabular}
    \par\smallskip
    \begin{flushleft}
        Notes. $^{\rm a}$ Columns 4 and 10 give the mass within a fiducial volume  $6'.67\times6'.67$ on the sky with depth $\pm20\ \lhmpc$. $^{\rm b}$ Columns 5 and 11 give the number of galaxies brighter than $24\ \rm mag$ in the ${\rm F444W}$ band that are associated with the FOF halo of each simulated cluster. The round brackets give the number of galaxies to this same limit within the fiducial volume.  $^{\dagger}$ The numbers for A2744 are based on \citet[]{Cha_Etal_2024}. The 2D projected mass $M_{\rm 2D}$ is obtained by integrating their convergence estimates over their lensing reconstruction field, and is compatible with their reported number of $\sim1.19\times 10^{15}\ {\rm M_{\odot}}$ within a $1\ {\rm Mpc}$ aperture. The member galaxy count denoted by $^{\ddagger}$ comes from their original member catalogue. The subsequent number in parentheses is the count used for this work, where we have cross-identified with the UNCONVER DR3 photometric catalogue \citep{Suess_Etal_2024} and eliminated galaxies with photometric redshifts inconsistent with A2744. This number should be compared with the numbers in parentheses for the simulated analogues. $^{\rm c}$ Columns 6 and 12 list the value of the relative mass ratio mismatch statistic $\mathcal{R}$ defined by Eq.~\ref{eq:rmrms} in Appendix~\ref{app:Procrustes}. This statistic is solely used as an extra criterion for identifying candidate analogues in \textbf{Group II}.
    \end{flushleft}
    \label{tab:DM_analogues}
\end{table*}

\subsubsection{Virtual JWST Images}
\label{sec:sam_gal_imag}
To generate virtual JWST images, we populate candidate halos with galaxies using the SAM models described in Section~\ref{sec:sam_model}. The SAM galaxies within the same rectangular box are then projected onto the fictitious observing plane along the same orientation used for the projected mass in Section~\ref{sec:kappa_maps} and also constrained to the field of $12'.5 \times 12'.5$. Virtual images are then generated following the procedure described in \citet[]{2013MNRAS.428..778O}. Each galaxy is assumed to comprise a disk and, optionally, a bulge component. Using the output information from the SAM code \texttt{L-GALAXIES}, the light distribution of each simulated galaxy is projected onto the presumed image plane. We summarise the key parameters used for simulating a galaxy in Appendix~\ref{app:sam_gal_params}.

In Figure~\ref{fig:mock_a2744_combo}, we present virtual images of all our A2744 analogues. The composite-colour images are overlaid with the corresponding projected mass distribution (yellow contours of the convergence $\kappa$ with a level spacing of 0.1 and smoothed with a Gaussian kernel with $\sigma=2''$; the lowest contour in each image is at $\kappa=0.2$). The composite RGB stacks utilise the F115W band image for blue, the F277W band for green, and the F444W band for red. These images are constructed so as to be directly comparable to Figure 4 of \cite{Cha_Etal_2024}; the overall morphology of the analogue clusters exhibits a remarkable similarity to that of A2744, as observed by JWST. 

Additionally, we construct maps of galaxy number density and F444W-band luminosity for these analogues. These maps are created by projecting their positions and F444W-band luminosities onto the image plane, maintaining the identical angular dimensions used for both the composite-colour image and the projected mass maps. As for the observational data, all galaxies brighter than magnitude 24 in the F444W band were used when making these images. These maps will be used for quantitative comparison with the real cluster. 

\section{Comparison with JWST UNCOVER Imaging and Lensing Analysis}
\subsection{The SL-WL Results of Cha et al. 2024}

\begin{figure}
    \includegraphics[width=0.47\textwidth]{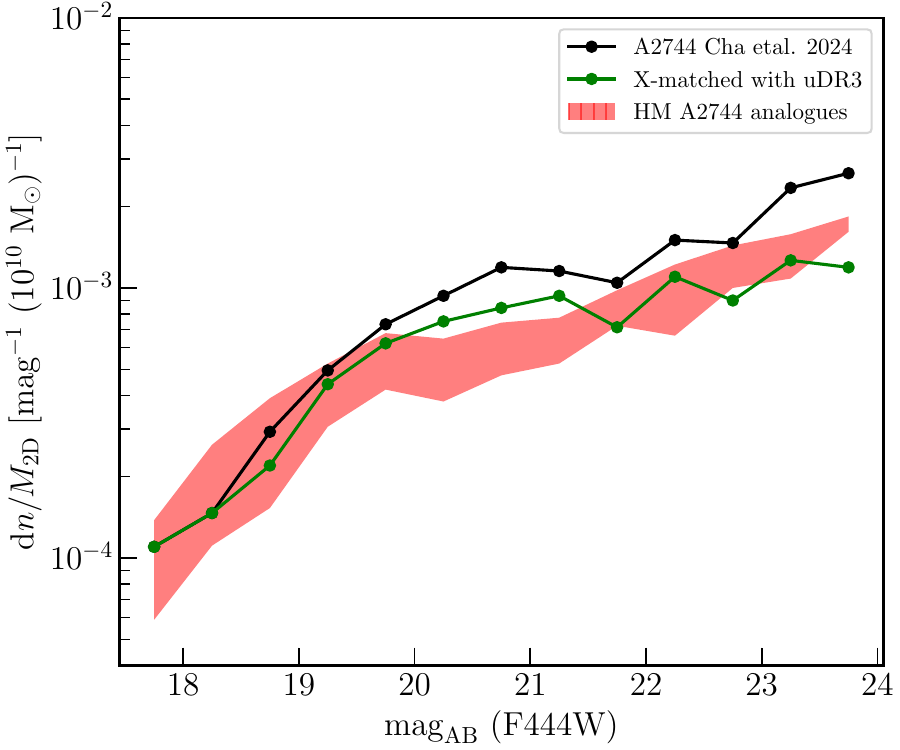}
    \caption{The luminosity function (LF) per unit projected mass for A2744 galaxies within the FOV analysed by \citet[]{Cha_Etal_2024}. Filled black dots connected with solid lines indicate the results when all ``non-background'' galaxies are used, while the green points and lines indicate the result after removing all galaxies with a spectroscopic or photometric redshift inconsistent with $z_{\rm cluster}=0.308$. The pink-shaded regions mark the area spanned by the projected, mass-normalised LFs of our nine simulated analogues of A2744.}
\label{fig:gal_LF}
\end{figure}

The UNCOVER imaging of A2744 \citep{UNCOVER_22} includes extremely deep observations of the cluster and its surrounding field using the Near Infrared Camera \citep[NIRCam;][]{2023PASP..135b8001R}. The $5\sigma$ limiting magnitudes are approximately 29.8 mag for point sources. These new, high-quality, deep and wide-field images highlight the power of JWST in revealing multiple-image systems and strongly lensed arcs within A2744, as well as reaching a very high number density of faint background galaxies for which shapes can be measured. 
The unprecedented amount of JWST data enables robust lensing reconstruction purely based on lensing data. Joint analysis of the strong and weak gravitational lensing signals across the full UNCOVER FOV guarantees the reconstruction of a mass density map of the cluster at remarkably high resolution. The map of \citet[]{Cha_Etal_2024} is based on $286$ SL multiple images, together with photo-$z$ selected background galaxies with a source density of $\sim 350\ {\rm arcmin^{-2}}$, and was constructed using the `MARS' algorithm \citep{Cha_Etal_2022}, a free-form, grid-based approach. It covers a FOV of 45 arcmin$^2$ with one arcsec$^2$ per pixel. 

It is important to note that the `MARS' approach does not rely on any form of light-traces-mass hypothesis when reconstructing the mass distribution of A2744. Indeed, it uses no information at all about the position and luminosity of cluster members. It thus gives a completely unbiased observational picture of the detailed relation between the lensing-derived mass distribution and the number/luminosity distributions of cluster member galaxies, providing a robust framework for studying how well our simulated clusters, based on a $\Lambda$CDM simulation and a semi-analytic galaxy formation model, agree with the observed relative distributions of galaxies and dark matter in a highly dynamic and asymmetric situation.

\begin{figure*}
    \includegraphics[width={0.65\textwidth}]{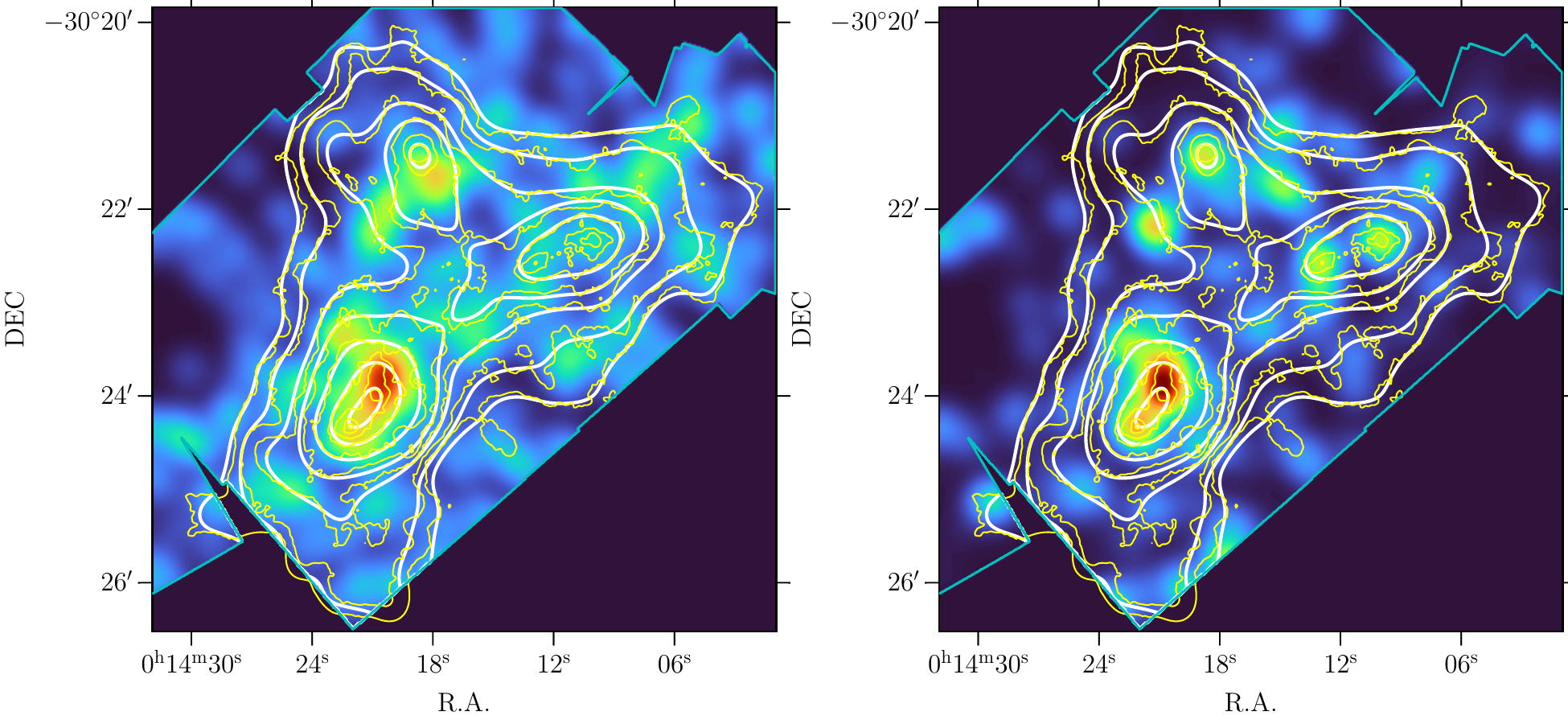}
    \includegraphics[width={0.33\textwidth}]{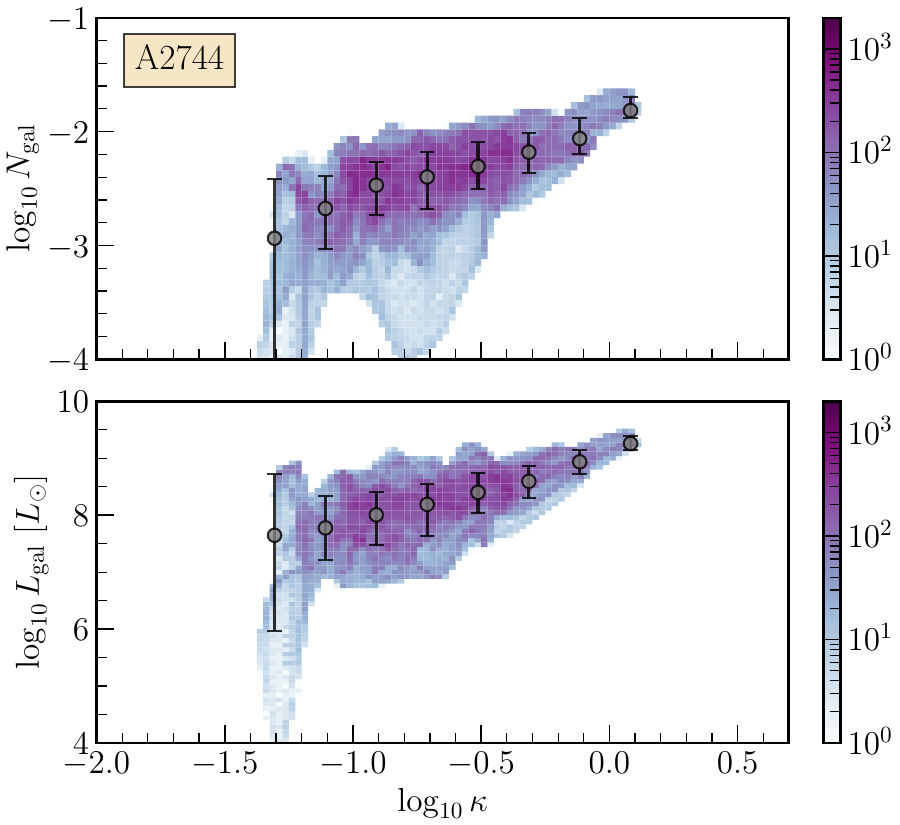}
    \caption{The left and middle panels present galaxy number and luminosity density maps for A2744 and are similar to \citet[]{Cha_Etal_2024}, but within the JWST footprint only and using only galaxies with redshift consistent with cluster membership. These maps have been smoothed with a Gaussian kernel with $\sigma=10^{\prime\prime}$ as described in the text. The dark blue regions are outside the JWST footprint, indicated by the cyan lines.  The fine yellow contours show the original map of the convergence $\kappa$ from \citet[]{Cha_Etal_2024} while the thicker white contours are obtained by smoothing this map in the same way as the underlying galaxy maps so that their resolutions match. In both maps contours are equally spaced in $\log_{10}\kappa$ with $\Delta\log_{10}\kappa=0.15$ and the lowest contour at $\kappa=0.15$. The right panels show two-dimensional histograms counting pixel numbers as a function of $\kappa$ and galaxy number density $N_{\rm gal}$ (top) or luminosity density $L_{\rm gal}$ (bottom). Open circles represent median values of the $y$-variable in equal-width bins of $\log\kappa$ with error bars indicating the 16th to 84th percentile range within each bin. The colour bars for the right panels encode pixel number. }
\label{fig:kappa_gal_n_l}
\end{figure*}

Before performing a detailed pixel-based comparison between the lensing-reconstructed mass map and the galaxy number/luminosity distributions, we re-examine the cluster membership criteria. In the original paper, the feature of the `MARS' algorithm being a free-form method does not require the positions of cluster member galaxies for reconstruction. Thus, the original catalogue focused on selecting robust background galaxies rather than strictly identifying cluster members. The potential member galaxies of A2744 were selected based on the F277W-F444W colour-magnitude relation. A very conservative colour window was taken to ensure that all cluster members and all foreground galaxies were excluded from the WL analysis. As a result, many of the excluded objects are not actually cluster members but rather background or foreground objects. 

To ensure our analysis is close to the original work of \citet[]{Cha_Etal_2024}, but nevertheless to exclude all objects that are definitely not part of the cluster, we use positional information to cross-identify galaxies from the original ``non-background'' catalogue with the newly publicly available DR3 catalogue from the UNCOVER team \citep{Suess_Etal_2024}. We then remove any galaxy that cannot be a cluster member using two redshift cuts. First, we eliminate galaxies with a spectroscopic redshift that differs by more than 0.02 from $z_{\rm cluster}=0.308$; for galaxies with no spectroscopic redshift, we eliminate those with photometric redshifts that differ by more than 0.09 ($\sim3\sigma$) from  $z_{\rm cluster}$. These are very conservative cuts. They yield a final sample of 504 galaxies brighter than $24\ \rm mag_{\rm F444W}$ within the FOV of the lensing reconstruction. 

In Figure \ref{fig:gal_LF} we compare the 2D-projected mass-normalised luminosity function (LF) of A2744 member galaxies in the lensing reconstruction field to similarly constructed mass-normalised LFs from our simulated analogues. For A2744, we show results both from the original catalogue of ``non-background'' galaxies (the black line with filled dots) and from the catalogue with non-members removed (the green line with filled dots). The region spanned by the mass-normalised LFs of our nine simulated analogues (\textbf{Group I} and \textbf{Group II} combined) is shown as the pink shaded region. The difference between the black and the green curves shows that our redshift cuts have removed galaxies at all magnitudes, but the correction is larger for fainter galaxies. After the removal of non-member galaxies, the mass-normalised luminosity function for A2744 agrees well with those predicted by our simulation, except possibly for a slight underprediction at intermediate magnitudes.

In Figure~\ref{fig:kappa_gal_n_l} we reproduce Figure 5 of \citet[]{Cha_Etal_2024} but using our cleaned-up member catalogue. We should note that although a $6'.67\times6'.67$ square field is presented, the UNCOVER imaging of A2744 only covers the highest density area of the cluster field. Galaxies are available only within the irregularly shaped JWST footprint (cyan lines), which occupies only half the square FOV. As we will show later, this introduces a bias in the pixel-based comparison between the observed and mocked A2744 clusters. Because no galaxies outside the JWST footprint could be used to produce the number and luminosity maps, image pixels near the footprint boundaries need special care in the smoothing process. We address this by constructing a binary mask matching the size and resolution of the galaxy number density $N_{\rm gal}$ and luminosity density $L_{\rm gal}$ maps. Pixels are set to unity within the JWST footprint and to zero outside it. After smoothing all maps with a Gaussian kernel with $\sigma=10^{\prime\prime}$, we correct boundary effects by dividing the smoothed $N_{\rm gal}$ and $L_{\rm gal}$ maps pixel-by-pixel by the smoothed mask map. For consistency, we also window and resmooth the $\kappa$-map with the same mask and the same kernel size of $10^{\prime\prime}$. In Figure~\ref{fig:kappa_gal_n_l}, only regions within the boundaries of the JWST footprint (the cyan lines) are displayed. The resmoothed $\kappa$-map is shown with heavy white contours, while the original $\kappa$-map from \citet[]{Cha_Etal_2024} is overlaid with finer yellow contours for comparison.

In the rightmost panels of Figure~\ref{fig:kappa_gal_n_l}, we present 2D histograms of pixel counts as a function of $\kappa$ and either galaxy number density $N_{\rm gal}$ (top) or galaxy luminosity density $L_{\rm gal}$ (bottom). In these plots, open circles represent the median values within equal-width bins of $\log_{10}\kappa$, with error bars indicating the region containing the central $68\%$ of the pixel count distributions. Both galaxy number and luminosity density exhibit a strong correlation with the dark matter density field, but with significant scatter. Qualitatively, this supports the light-traces-mass assumption widely adopted in many mass reconstruction models.

\subsection{The A2744 analogues in the HM simulation}
\begin{figure*}
    \includegraphics[width={0.63\textwidth}]{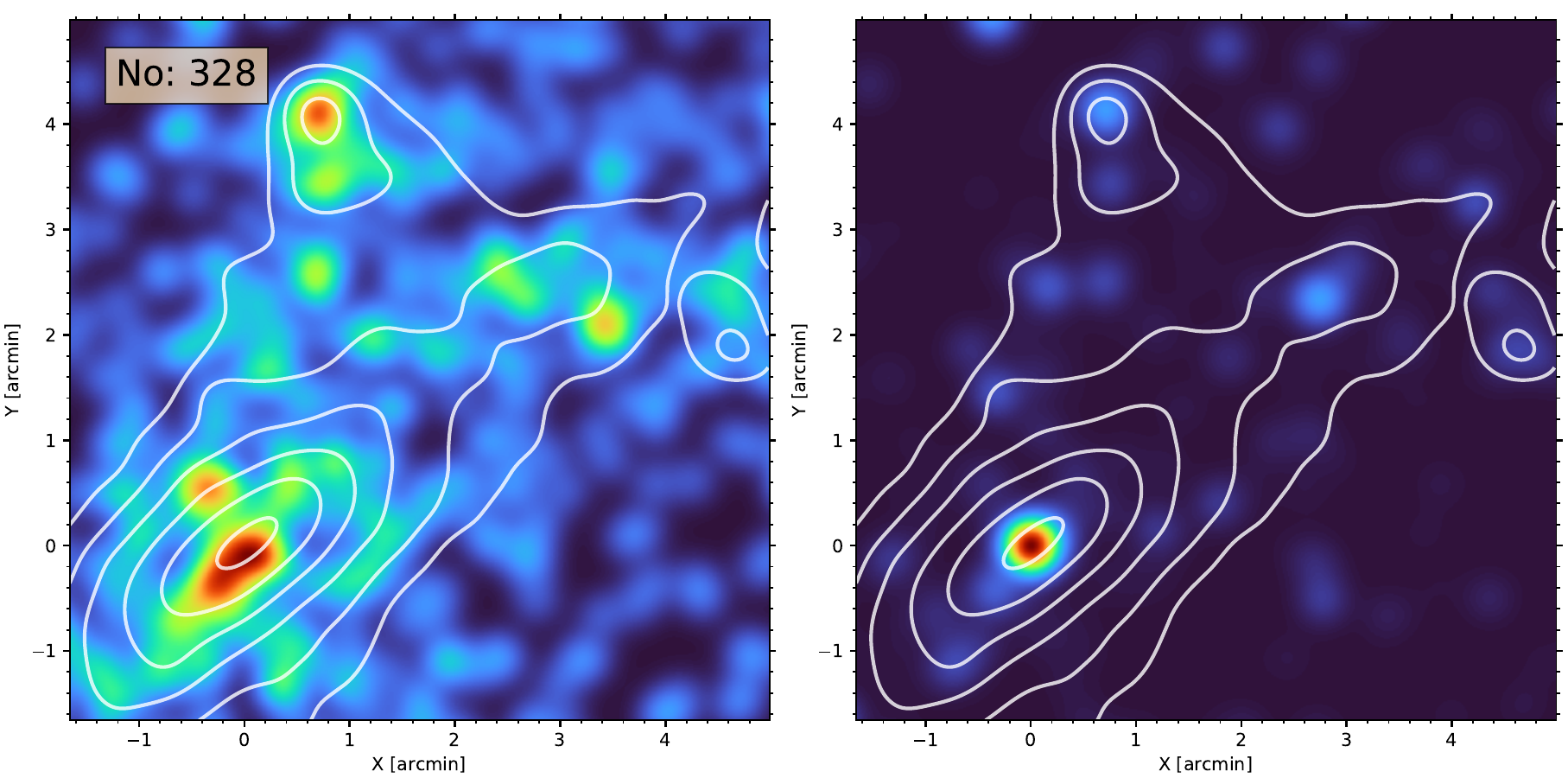}
    \includegraphics[width={0.34\textwidth}]{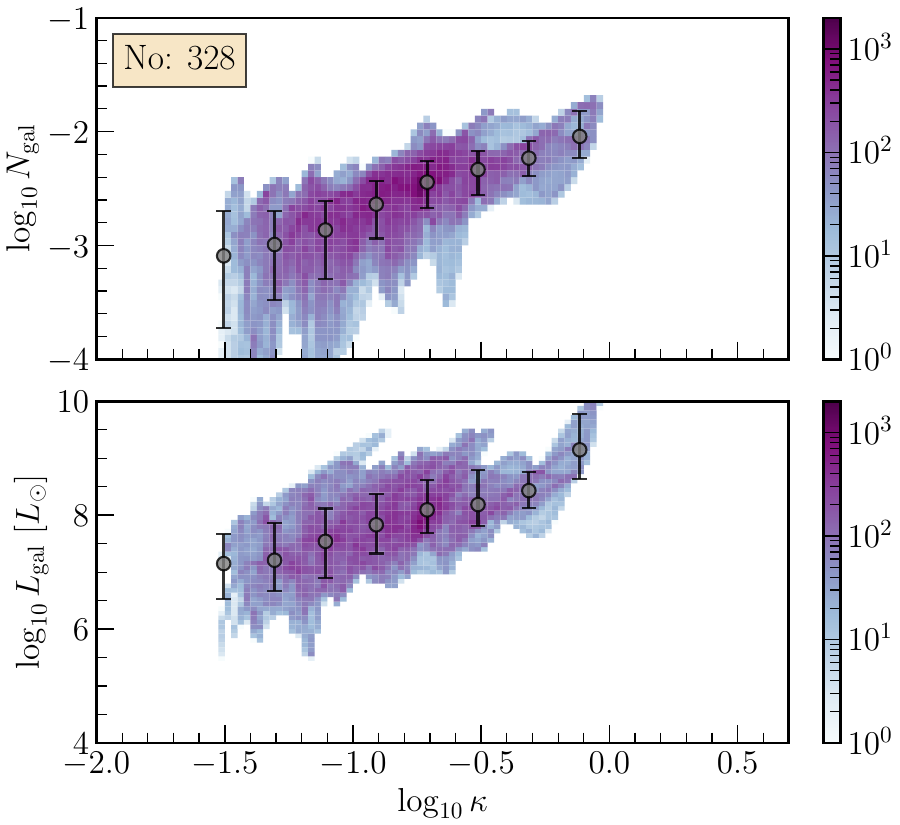}

    \includegraphics[width={0.63\textwidth}]{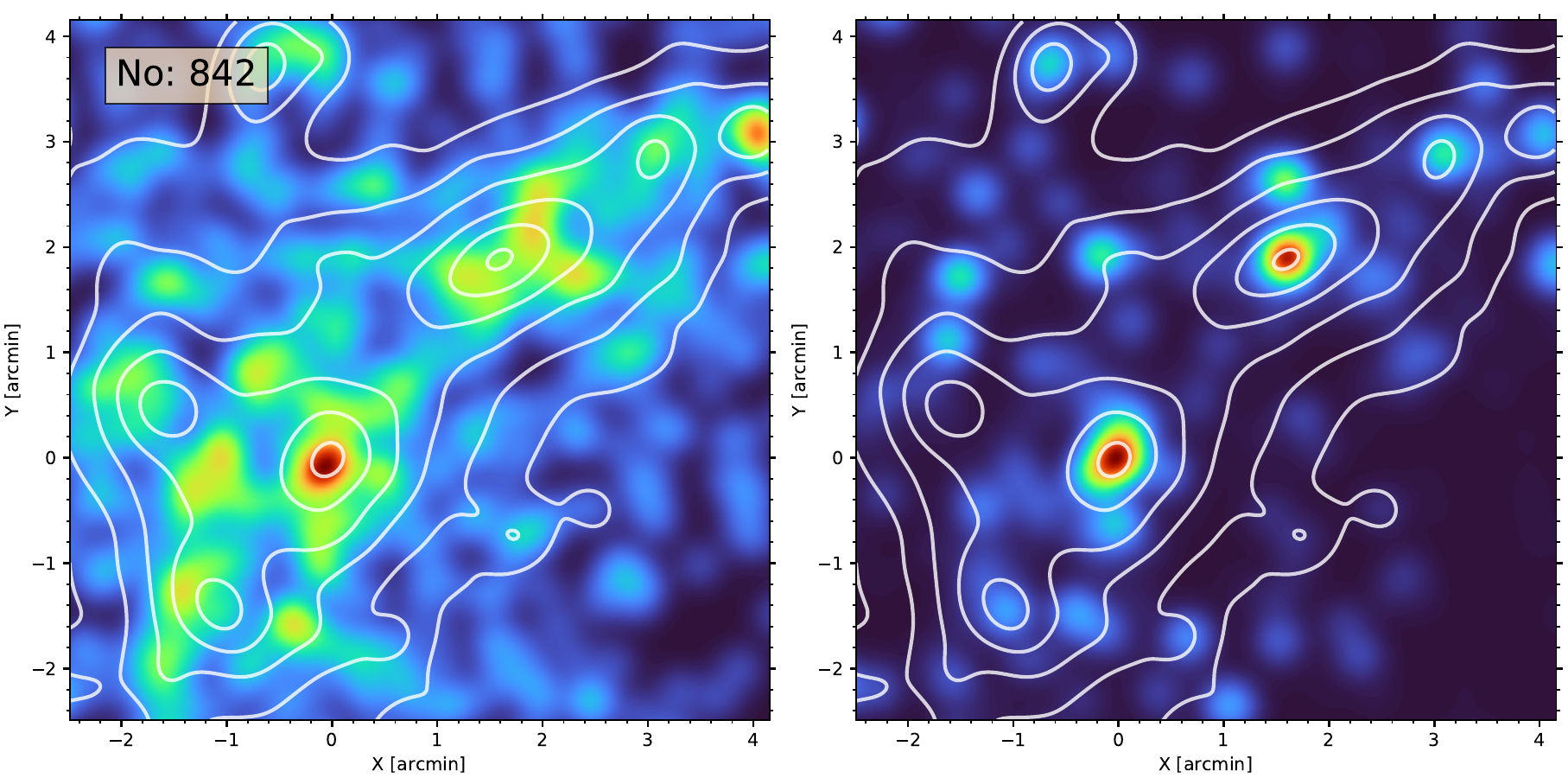}
    \includegraphics[width={0.34\textwidth}]{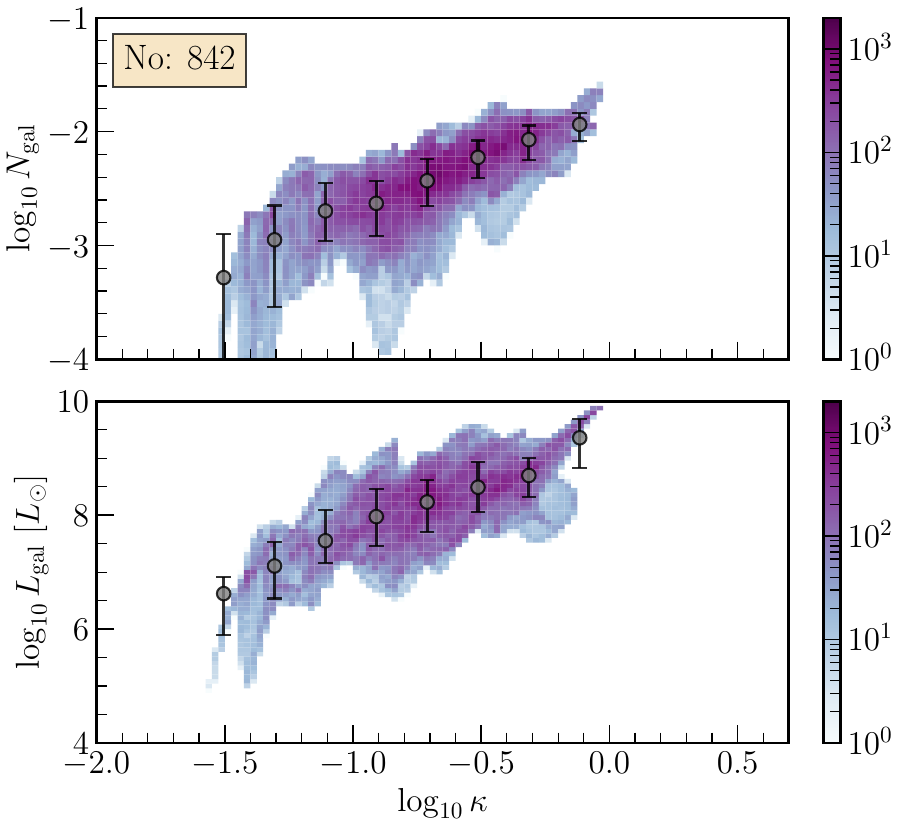}
    
    \caption{Selected A2744 analogues in the HM simulation (Nos. 328 and 842). The colour scales and contour values are identical to those in Figure~\ref{fig:kappa_gal_n_l}. Again, the left and middle panels show the galaxy number density and luminosity density maps, respectively, while the right panels show two-dimensional histograms counting pixel numbers as a function of $\kappa$ and galaxy number density $N_{\rm gal}$ (top) or luminosity density $L_{\rm gal}$ (bottom).}
\label{fig:HM_A2744_samplesMain}
\end{figure*}

We now make a quantitative comparison of all nine of our analogue clusters (see Figure~\ref{fig:mock_a2744_combo}) with the observed A2744 system. We restrict the field of view (FOV) to $6'.67\times6'.67$ in order to match the area covered by the lensing reconstruction. In Table~\ref {tab:DM_analogues}, we present member galaxy counts associated with each FOF cluster, along with total galaxy counts (brighter than 24 mag. in the F444W band) within $\pm 20\ \lhmpc$ in projection within this $6'.67\times6'.67$ FOV. These counts range from 484 to 807 across our analogue sample, consistent with the observed count of 504 galaxies. The total projected masses for these analogue clusters range from $1.02\times10^{15}\ {\rm M_{\odot}}$ to $1.61\times10^{15}\ {\rm M_{\odot}}$, also bracketing the observationally derived mass estimate of $1.19\times10^{15}\ {\rm M_{\odot}}$ reported by \citet[]{Cha_Etal_2024}.  For the five {\bf Group I} clusters, the mean and scatter in the number of galaxies per $10^{14}M_\odot$ is $47.1\pm2.1$, which agrees essentially perfectly with the observed value of 46.2 in A2744. For the four {\bf Group II} clusters, the value is slightly higher, $51.3\pm1.5$, reflecting the fact, already noted above, that these are line-of-sight projections of several adjacent but dynamically connected clumps of somewhat lower mass.

Figure~\ref{fig:HM_A2744_samplesMain}  shows, for two examples (Nos. 328 and 842), the galaxy number $N_{\rm gal}$ and luminosity $L_{\rm gal}$ maps and the 2D histograms of pixel counts as a function of $\log_{10}\kappa$ and either $N_{\rm gal}$ (upper) or $L_{\rm gal}$ (lower). These can be compared directly with Figure~\ref{fig:kappa_gal_n_l}. Visually, both analogues appear very similar to A2744, sharing features such as multiple clumps, in which the galaxy number and luminosity distributions closely trace the underlying mass distribution.  For both analogues, the mean relation and the scatter in both 2D histograms are similar to those for A2744. However, the slopes of the simulated relations are slightly steeper than those observed. The same is true for the remaining analogues. For completeness, we present the corresponding plots for all other analogues in Appendix~\ref{app:gal_nl_rest}. It seems likely that this difference reflects a bias caused by the fact that the region targeted for observation in A2744 was chosen specifically to cover the regions of highest galaxy surface density. We come back to this below.

\subsection{The pixel-based comparison between the observed and mocked A2744 clusters}
\begin{figure*}
    \includegraphics[width={0.475\textwidth}]{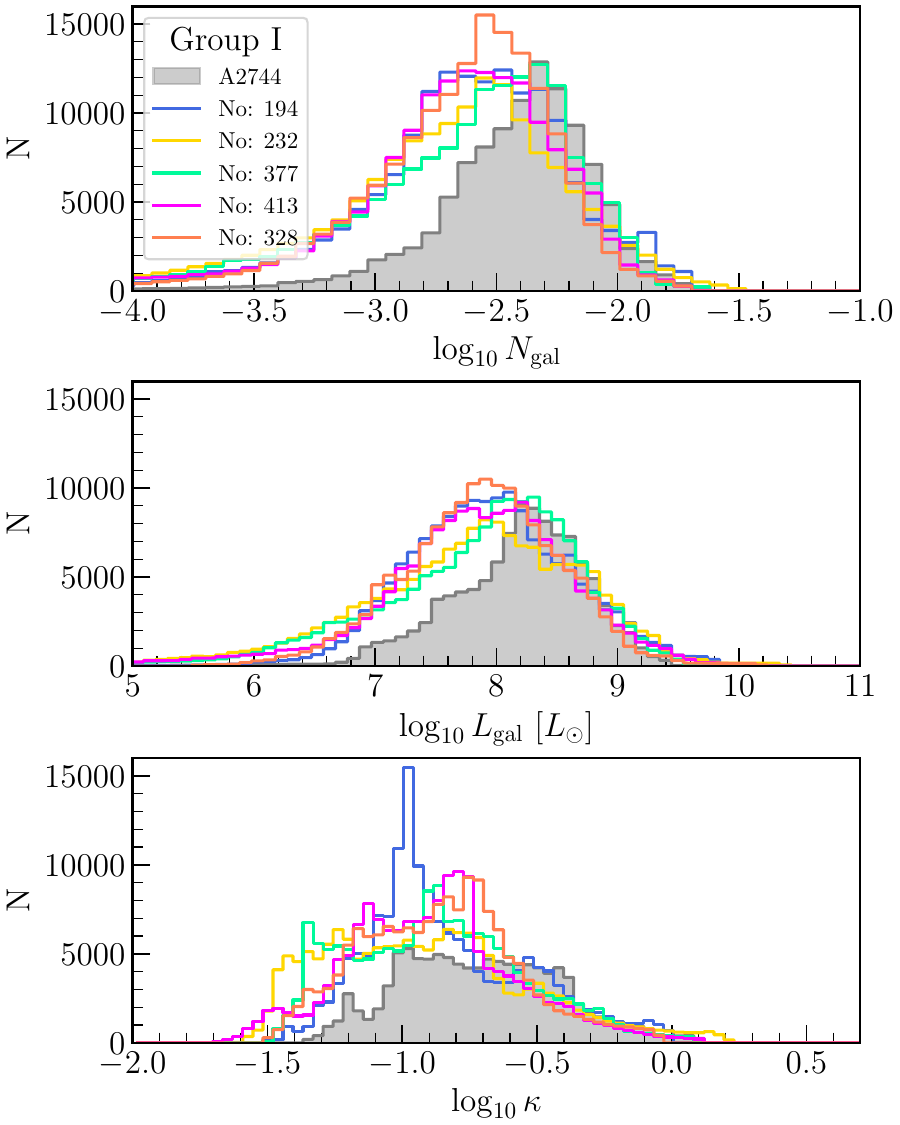}
    \includegraphics[width={0.475\textwidth}]{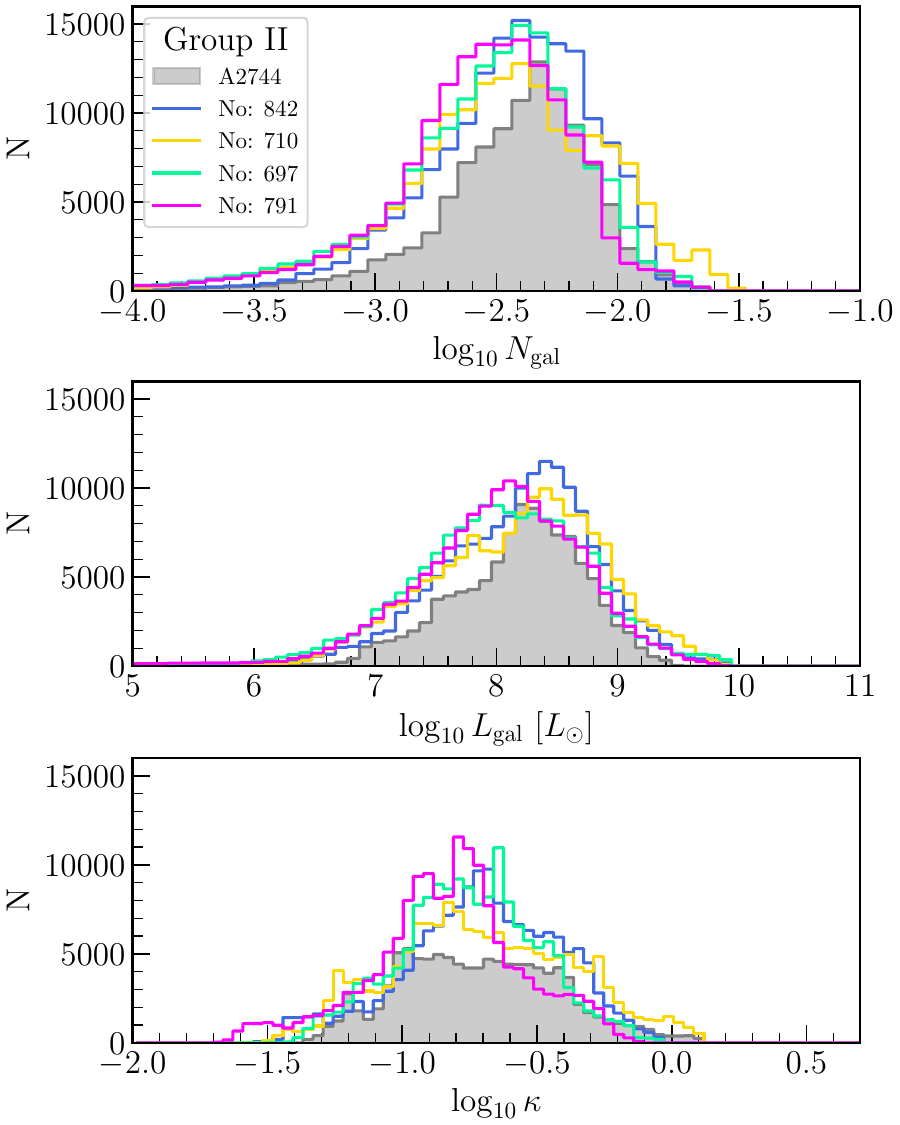}
    \caption{Histograms of galaxy number density $N_{\rm gal}$ (top), galaxy luminosity density $L_{\rm gal}$ (middle), and convergence $\kappa$ (bottom) comparing pixel counts within a FOV of $6'.67 \times 6'.67$ for our A2744 analogues (coloured solid lines) with counts within the JWST footprint for the observed cluster (hatched grey histogram; from \citet[]{Cha_Etal_2024} after resmoothing to $\sigma=10^{\prime\prime}$). The left and right panels are for clusters in Group I and Group II, respectively. Note that the JWST footprint covers only about half of the full square FOV, explaining why there are fewer pixels in the histograms for the observed cluster than in those for its analogues.}
    \label{fig:gLumNumHist1D}
\end{figure*}

In this subsection, we perform a pixel-based comparison between our A2744 analogues and the observed cluster. Figure~\ref{fig:gLumNumHist1D} shows histograms of pixel values for galaxy number density $N_{\rm gal}$, galaxy luminosity density $L_{\rm gal}$, and convergence $\kappa$. All three maps were smoothed with a Gaussian kernel of dispersion $10^{\prime\prime}$ before compiling these histograms. We note that while we use a $6'.67\times6'.67$ square FOV for the A2744 analogues, data are only available for the real cluster within the irregularly shaped JWST footprint. This covers just over half of the square FOV and was clearly chosen to cover the highest density area of the real cluster. This bias is obvious in Figure~\ref{fig:gLumNumHist1D}, where the total number of pixels in each histogram is about a factor of two smaller for A2744 than for the simulated analogues; the histograms for all three quantities coincide at the high density end, however, showing that the highest density regions of the simulated clusters agree very well with the observed object.

\begin{figure*}
    \includegraphics[width=0.9\textwidth]{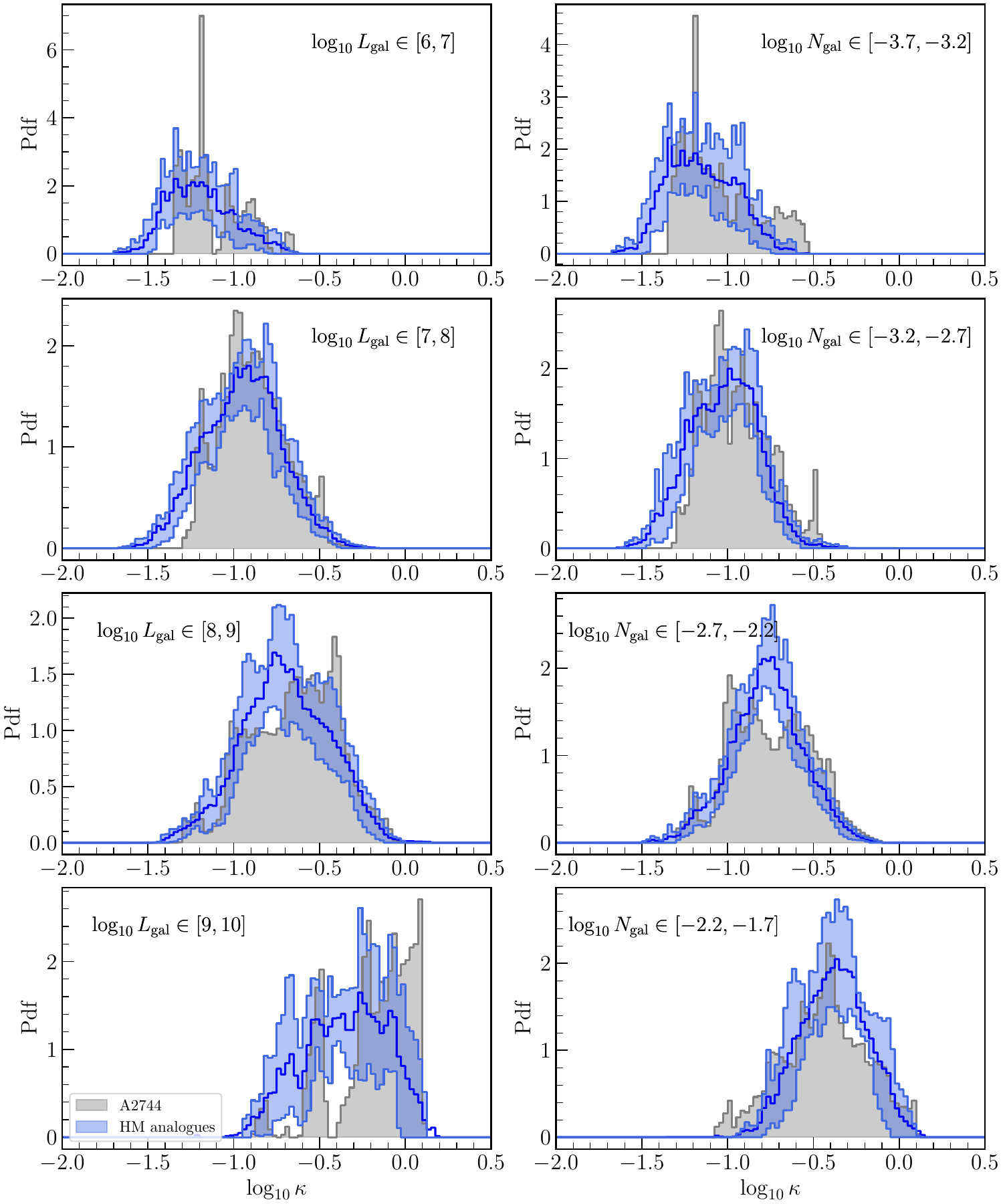}
    \caption{Normalised distributions of convergence (resmoothed to $\sigma=10^{\prime\prime}$) for pixels which have galaxy luminosity density $L_{\rm gal}$ (left) or galaxy number density $N_{\rm gal}$ (right) in the intervals specified by the legend in each panel. Results for A2744 are shown by the hatched grey histogram, while the solid blue line and the hatched blue region around it show the mean and the 16 to 84 percentile range of the distributions for the nine simulated analogues.}
    \label{fig:KappaBinedPdfs}
\end{figure*}

In order to compare simulation and observation in a way that is almost independent of this bias, we study the distribution of $\kappa$ in fixed intervals of  $L_{\rm gal}$ and $N_{\rm gal}$. Specifically,  Figure~\ref{fig:KappaBinedPdfs} shows the distributions of $\kappa$ for four equal intervals of $\log L_{\rm gal}$ spanning the range from $6$ to $10$, and for four equal intervals of $\log N_{\rm gal}$ spanning the range from $-3.7$ to $-1.7$. In this figure, all histograms are normalised to unit integral. The distribution of $\kappa$-values is very similar between A2744 and its simulated analogues in all panels. Thus, once the observational bias in the selection of the JWST fields for A2744 is taken into account, the detailed agreement between simulation and observation is excellent. 

Our detailed, pixel-based comparison between the simulated A2744 analogues and the JWST observations demonstrates that such a rare and complex object can arise in a $\Lambda{\rm CDM}$ universe. The SAM plays an important role in the "mass–light" comparison. The SAM model prescriptions of \citet{2015MNRAS.451.2663H} and \citet{Pei_etal_24} implemented in this work was not re-tuned specifically for the HM simulation. The model we use here has previously been shown to successfully reproduce several independent relations directly relevant to this study, such as the stellar-to-halo mass relation and the statistics of the spatial distribution of galaxies. The excellent match we find for the luminosity function (as shown in Figure \ref{fig:gal_LF}) also provides a posteriori support for this model in this extreme cluster regime.

In order to test whether a $\Lambda{\rm CDM}$ universe can produce systems like A2744, we adopt a previously calibrated and observationally validated SAM as our baseline, since exploring variations in model assumptions/parameters without simultaneously ensuring consistency with previously used observational constraints would likely lead to models that no longer correctly describe the general galaxy population. This would compromise our test of whether the model's linking of light to mass extends successfully to this extreme case.  Future studies employing alternative SAMs or full hydrodynamical simulations could test the robustness of these results and enable a more comprehensive comparison.

\section{Conclusion}
In this paper, we have presented the HyperMillennium (HM) simulation, an extremely large cosmological simulation designed to meet the demands of next-generation galaxy surveys. The HM simulation is performed with a new version of the \texttt{PhotoNs-3.7} code. It evolves approximately 4.2 trillion dark matter particles within a cubic volume of (2.5 $h^{-1}$ Gpc)$^3$, achieving a mass resolution of $3.216 \times 10^8 h^{-1}M_\odot$ and a force resolution of 3.0 $h^{-1}$ kpc. 

The HM simulation outputs 100 snapshots across the redshift range from 20.4 to 0. The total disk requirement for long-term data storage is approximately 12 PB. Its combination of large volume and high resolution makes it particularly well-suited for studying large-scale structures and rare object populations. At $z=0$, the simulation contains approximately $8 \times 10^9$ FOF groups, and of these, 3,740 are more massive than $10^{15}\ \msun$. The most massive group has a FOF mass of $4.9 \times 10^{15}\ \msun$.  Subhalo catalogues were generated at all output times with the SUBFIND algorithm, and merger trees were constructed following the methods described in \citet{Angulo12} and \citet{Angulo14}. In principle, these allow semi-analytic galaxy populations to be generated for all objects and at all times using methods similar to those of \cite{Guo_11} or \citet{2015MNRAS.451.2663H}, although this becomes computationally demanding if galaxy catalogues are required for the entire volume. In practice, it is often sufficient, as in this paper, to make catalogues for specific objects or subregions.

In this first paper of the HM project, we take advantage of the large volume and high mass resolution to focus on the most massive structures and investigate whether the dark and luminous structure of galaxy clusters in the extreme high mass tail, exemplified by Abell 2744 (A2744), is reproduced in detail by the simulation and by semi-analytic galaxy catalogues built on top of it. Analysing the snapshot at redshift $z=0.2995$, closest to that of A2744 ($z_{\rm cluster}=0.308$), we carry out a \texttt{Procrustes} analysis of the  851 systems with $M_{\rm FOF} > 10^{15}\ \msun$. This geometric morphometry technique identifies the optimal simulated halos and the optimal projection view to obtain objects with five prominent mass structures resembling A2744.

For candidate A2744 analogues identified in this way, we generate galaxies to populate the clusters and their immediate foreground and background regions using the SAM code \texttt{L-GALAXIES} \citep{MS_06}, updated to incorporate model prescriptions from \citet{2015MNRAS.451.2663H} and \citet{Pei_etal_24}. We then generate JWST/NIRCam-like images and lensing convergence maps, and we compare these with the observed maps for A2744 as reported by \citet[]{Cha_Etal_2024}. 

This procedure allows us to make a detailed apples-to-apples comparison of the joint distribution of light and mass in A2744 to those in its simulated analogues. Specifically, we compare pixel-based statistics for the joint distributions of galaxy number density $N_{\rm gal}$, galaxy luminosity density $L_{\rm gal}$, and projected mass density $\kappa$. These comparisons demonstrate that the selection of JWST fields for A2744 was strongly biased towards the regions of highest galaxy surface density.  Once this is taken into account, the distributions in the real cluster and in its analogues are in excellent agreement. This demonstrates that even in the highly nonlinear and irregular regions associated with the extreme tail of the galaxy cluster population, the joint distribution of projected mass density and projected galaxy number or luminosity density agrees between this $\Lambda{\rm CDM}$ simulation and observation down to scales of order 50~kpc. 

Our HM simulation, combined with a previously calibrated SA model, reproduces successfully and in detail the observed properties of the extreme cluster A2744, confirming the validity both of the SAM and of the underlying $\Lambda{\rm CDM}$ model in this extreme environment, and giving further confidence in the power and fidelity of our 'simulation $+$ SAM' approach. This is able to produce similar quality galaxy and mass catalogues out to high redshift and over the HM's entire comoving volume of 50.4 Gpc$^3$.

\section*{Acknowledgements}
This work is supported by the Strategic Priority Research Program of the Chinese Academy of Sciences (Grant No. XDB0500203), the National Key Research and Development Program of China (Grant No. 2023YFB3002501), Ministry of Science and Technology of China (Grant No. 2020SKA0110100, 2020SKA0110401, 2022SKA0110201), and the National Natural Science Foundation of China (NSFC) (Nos. 12033008, 12588202, 12425303).
ML also acknowledges the support from the National Key Research and Development Program of China (No. 2022YFA1602903).
CSF was supported by the European Research Council (ERC) Advanced Investigator grant DMIDAS (GA 786910) and the STFC Consolidated Grant ST/T000244/1.
SC acknowledges support for the current research from the National Research Foundation of Korea under the programs 2022R1A2C1003130, RS-2023-00219959, and RS-2024-00413036. 
REA was supported by the Spanish Ministry of Science, Innovation and Universities through grant number PID2024-161003NB-I00.

The numerical calculations in this study were carried out on the ORISE Supercomputer.


\section*{Data Availability}

The HyperMillennium simulation is scheduled for full public release in 2026. The underlying data of this article will be made available to interested readers following a reasonable request addressed to the corresponding authors.

\bibliographystyle{mnras}
\bibliography{HM-P1-JWST} 




\appendix

\section{The basis of the Procrustes analysis}
\label{app:Procrustes}
In Section~\ref{sec:proc_analysis}, we present the basic ideas to implement the \texttt{Procrustes} analysis in our A2744 analogue selections. With the observed positions of the five A2744 bright galaxies, $\mathbf{P}=\left\{\mathbf{p}_1, \mathbf{p}_2, \ldots, \mathbf{p}_5\right\}$, and the 2D projected positions of the five most massive substructures of the halo under consideration, $\mathbf{Q}=\left\{\mathbf{q}_1, \mathbf{q}_2, \ldots, \mathbf{q}_5\right\}$, the \texttt{Procrustes} analysis seeks the optimal combination of rotation ($\mathbf{R}$), translation ($\mathbf{t}$), and uniform scaling ($\mathbf{s}$) that minimises the sum of squared distances $D_{\rm proc}$. 

In this work, we present two A2744 analogue samples: one consisting of candidates purely determined by geometrical similarity and the other consisting of candidates selected with additional mass information. The details are introduced in the following sections.

\subsection{The Pure Geometrical Selection}
\label{app:PGrp1}
To ensure robust candidate selection, we implement an iterative refinement process. For each halo, we generate multiple random 3D rotations followed by projections, after which the \texttt{Procrustes} distance $D_{\rm proc}$ is computed for each orientation. To determine the optimal alignment, we minimise the distance $D_{\rm proc, min}$ to find its minimum value using the \texttt{scipy.optimize} routine \citep[SciPy:][]{Scipy_1.17}. The projection yielding $D_{\rm proc, min}$ is then retained as the best match.

To assess the stability of \texttt{Procrustes} results, we implement two optimization termination criteria, one is with a fixed iterations limit and the other is automatic termination. When using a fixed number, 100 iterations are sufficient to achieve stable results. Alternatively, \texttt{scipy.optimize.minimize} routine using the \texttt{L-BFGS-B} regulation typically converges on $D_{\rm proc, min}$ within a few tens of iterations, offering greater computational efficiency. The two strategies yielded nearly identical alignment results. Combined with visual inspection, our final selection of candidates is robust to the specific optimisation strategy employed.

After the optimisation process, all halos are ranked according to their $D_{\rm proc, min}$, with smaller values signifying closer morphological resemblance to A2744. We finally select candidates with $D_{\rm proc, min}$ falling within the lowest 10th percentile of the distribution. We complement this procedure with careful visual examination of the projected cluster configurations. The resulting set, identified as halos Nos. 194, 232, 377, 413, 328, forms \textbf{Group I}—the geometrically selected A2744 analogues.

\subsection{The Mass Refinement of Analogue Identification}
\label{app:PGrp2}
While the approach above ensures geometric similarity, it does not incorporate information about the dark matter mass of member galaxies. However, strong- and weak-lensing studies provide mass estimates for the prominent galaxies in A2744, for example, \citet[]{Cha_Etal_2024} have $(2.084, 1.947, 1.052, 1.794, 1.392)\times 10^{14}\ {\rm M_{\odot}}$ of their WL+SL NFW profile-fitting results for their quoted (BCG-N, BCG-S, G1, G2, and G3); \citet[]{Furtak_Etal_2023} have $(8.52, 4.98, 4.36, 3.92, 3.56)\times 10^{12}\ {\rm M_{\odot}}$ of their quoted BCGs (BCG2, BCG1, NW-BCG2, NW-BCG1, and N-BCG) for their SL model of A2744.

For this reason, we consider introducing a mass ratio criterion. For A2744, we take $M_{\mathbf{p}_i}$ to be the mass of the five BCGs derived from the WL+SL NFW fits in \citet[]{Cha_Etal_2024}. For simulated systems, we define $M_{\mathbf{q}_i}$ as the subhalo masses of the five most massive structures of each analogue. We then normalised both $M_{\mathbf{p}_i}$ and $M_{\mathbf{q}_i}$ by the total mass of their respective five most massive structures, forming a set of relative masses for A2744, and for each analogue halo accordingly, as
\begin{equation}
    \widetilde{M}_{\mathbf{p}_i}=\frac{M_{\mathbf{p}_i}}{\sum_{i=1}^{5} M_{\mathbf{p}_i}}, \ \widetilde{M}_{\mathbf{q}_i}=\frac{M_{\mathbf{q}_i}}{\sum_{i=1}^{5} M_{\mathbf{q}_i}}.
\end{equation}
Then we define a relative mass ratio mismatch statistic, $\mathcal{R}$, as
\begin{equation}
    \mathcal{R} = \frac{1}{N}\sum_{i}^{N} \bigg[\frac{2(\widetilde{M}_{\mathbf{p}_i}-\widetilde{M}_{\mathbf{q}_i})}{\widetilde{M}_{\mathbf{p}_i}+\widetilde{M}_{\mathbf{q}_i}}\bigg]^2 , \ N=5.
    \label{eq:rmrms}
\end{equation}
The value of $\mathcal{R}$ is bounded between 0 and 4. A value of $\mathcal{R}=0$ indicates a perfect match where the relative mass distributions are identical, while $\mathcal{R}=4$ represents the worst-case scenario with maximal discrepancy. Therefore, a lower value of $\mathcal{R}$ indicates a better match in the relative mass ratio of the top five substructures, and this statistic could serve as an extra criterion for identifying the most realistic analogues in the simulation. 

For candidate selection, we first rank all halos in ascending order of their $\mathcal{R}$ values. The $\mathcal{R}$ values for our halo sample range from 0.39 to 3.32. We then perform the same \texttt{Procrustes} analysis described in Appendix \ref{app:PGrp1}. The final selection, combined with visual inspections, identifies four additional A2744 analogues: Nos. 842, 710, 697, and 791 are designated as \textbf{Group II}.

\section{SAM galaxies parameters for Constructing Virtual JWST Images}
\label{app:sam_gal_params}
We created mock images of the simulated cluster member galaxies following the procedure described in \citet[]{2013MNRAS.428..778O}. Below, we summarise the key parameters for a simulated galaxy used in the procedure:

\begin{itemize}
    \item \textit{$x_c, y_c$}: The galaxy's position relative to the centre of the FOV on the image plane, expressed in pixel coordinates.

    \item mag: The total apparent (AB) magnitude of the galaxy in the chosen filter. We output magnitude values for eight NIRCam bands, F090W, F115W, F150W, F200W, F277W, F356W, F410M, and F444W. The filter properties used in \texttt{L-GALAXIES} for these bands are taken from \texttt{The SVO Filter Profile Service}\footnote{\url{https://svo2.cab.inta-csic.es/theory/fps3/index.php?mode=browse&gname=JWST&gname2=NIRCam&asttype=}}.

    \item \textit{B/T}: The bulge-to-total ratio, used to assign magnitudes to the bulge and disk components via:
    $$
    {\rm mag}_{\rm b} = {\rm mag} - 2.5 \log_{10}(B/T), \quad
    {\rm mag}_{\rm d} = {\rm mag} - 2.5 \log_{10}(1-B/T).
    $$

    \item \textit{$R_0$}: The scale height of the disk defined as \texttt{StellarDiskRadius}/3, or the bulge's half-light radius defined as \texttt{BulgeSize}. Both \texttt{StellarDiskRadius} and \texttt{BulgeSize} are taken from the SAM (semi-analytic model) data.

    \item \textit{$D_{\rm A}$}: The angular diameter distance to redshift $z=0.308$, the redshift of A2744. This is used to transform the physical sizes of galaxies into angular sizes in arcseconds.

    \item \textit{$\theta, \cos(\phi)$}: The position angle and aspect ratio after projection. For simplicity, we assign the aspect ratio $\cos(\phi)$ for disk galaxies using the \texttt{CosInclination} value from the SAM. The position angles ($\theta$) are generated randomly with a uniform distribution. For the bulge, we assumed it is spherical and set $\cos(\phi)=1, \theta=0$.
\end{itemize}

Using these parameters, the observed flux in a JWST/NIRCam band of each galaxy is related to the AB magnitude via
\begin{equation}
    F_{u} = 10^{-0.4 (m_{\rm AB} - m_{\rm ZP})},
\end{equation}
where we adopt a zero point of $m_{\rm ZP} = 28.9$, corresponding to a flux unit of 10 nJy. 

The brightness profile of the galaxy at the pixel position $(x^{\prime}, y^{\prime})$ on the image plane is modelled using a Sersic profile:
\begin{equation}
    I(r) \sim F_{u}\ e^{-(r/R_0)^{1/n}}, 
\end{equation}
the Sersic index is $n=1.5$ for the disk and $n=3.5$ for the bulge component. Here, $R_0$ represents the scale height of the disk or the bulge's half-light radius, while $r = \sqrt{x^{\prime 2} + y^{\prime 2}}$ is the distance to the central pixel of the galaxy, with $x^{\prime} = x - x_c$ and $y^{\prime} = y - y_c$.

\section{Galaxy number density and luminosity maps for other A2744 analogous}
\label{app:gal_nl_rest}
In Figures~\ref{fig:HM_A2744_samples_G1} and \ref{fig:HM_A2744_samples_G2}, we present the pixel-level comparisons of galaxy number and luminosity maps as a function of $\log_{10}\kappa$, for the remaining analogues that are not illustrated in Figure~\ref{fig:HM_A2744_samplesMain}. 

\begin{figure*}
    \includegraphics[width={0.63\textwidth}]{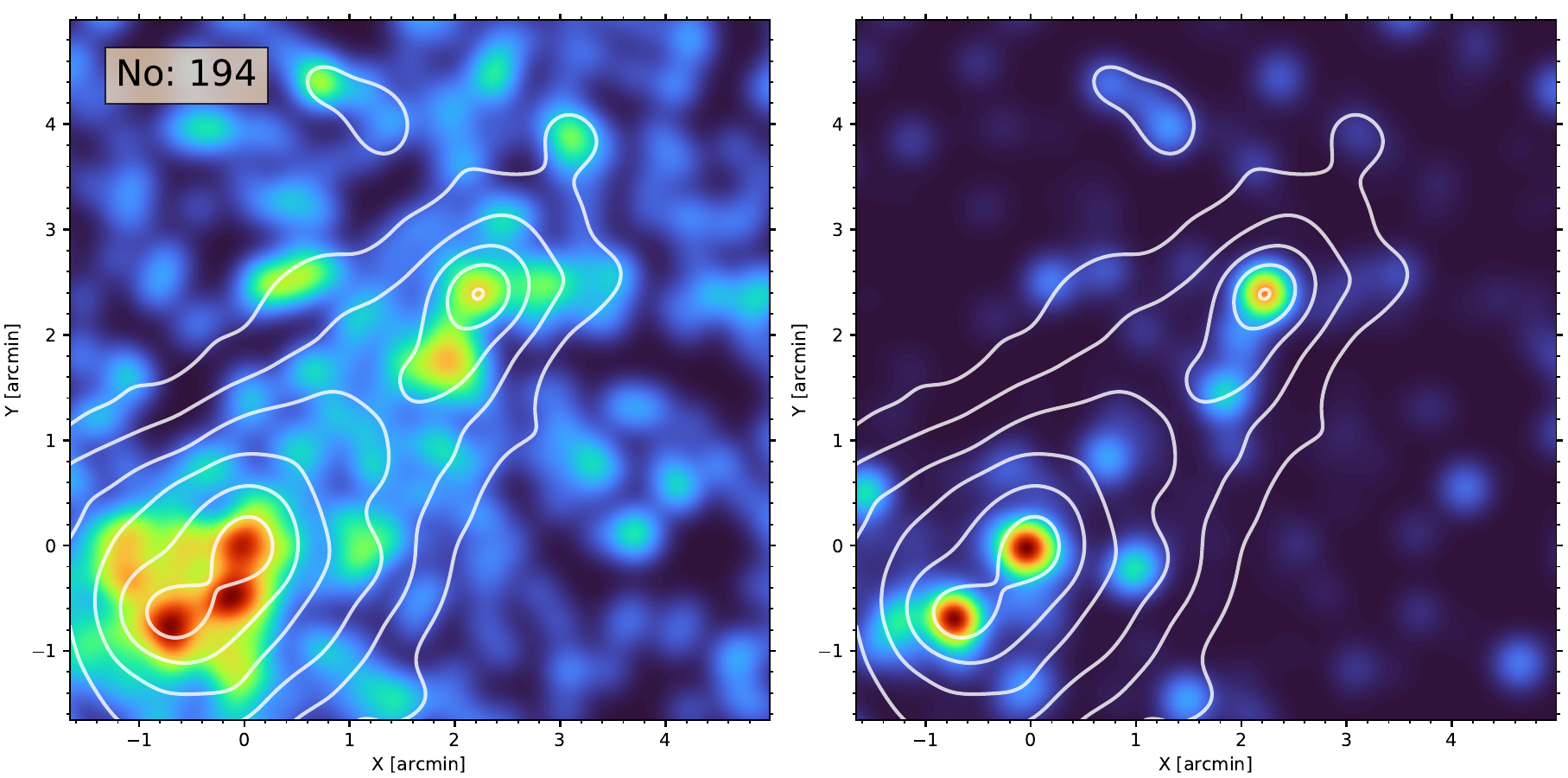}
    \includegraphics[width={0.34\textwidth}]{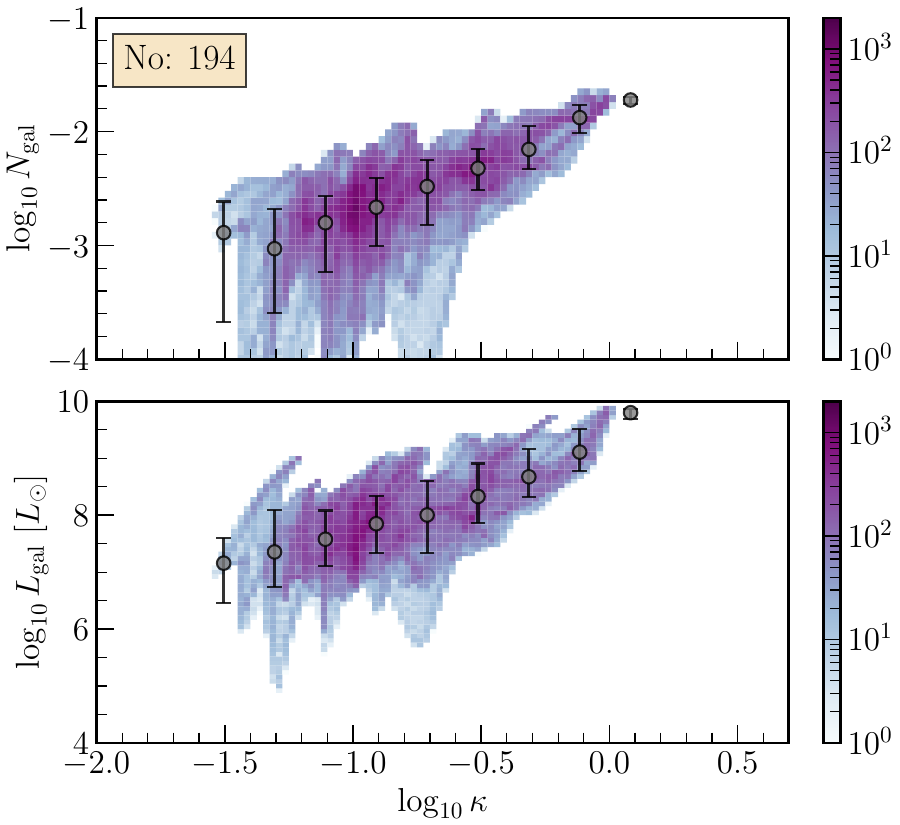}

    \includegraphics[width={0.63\textwidth}]{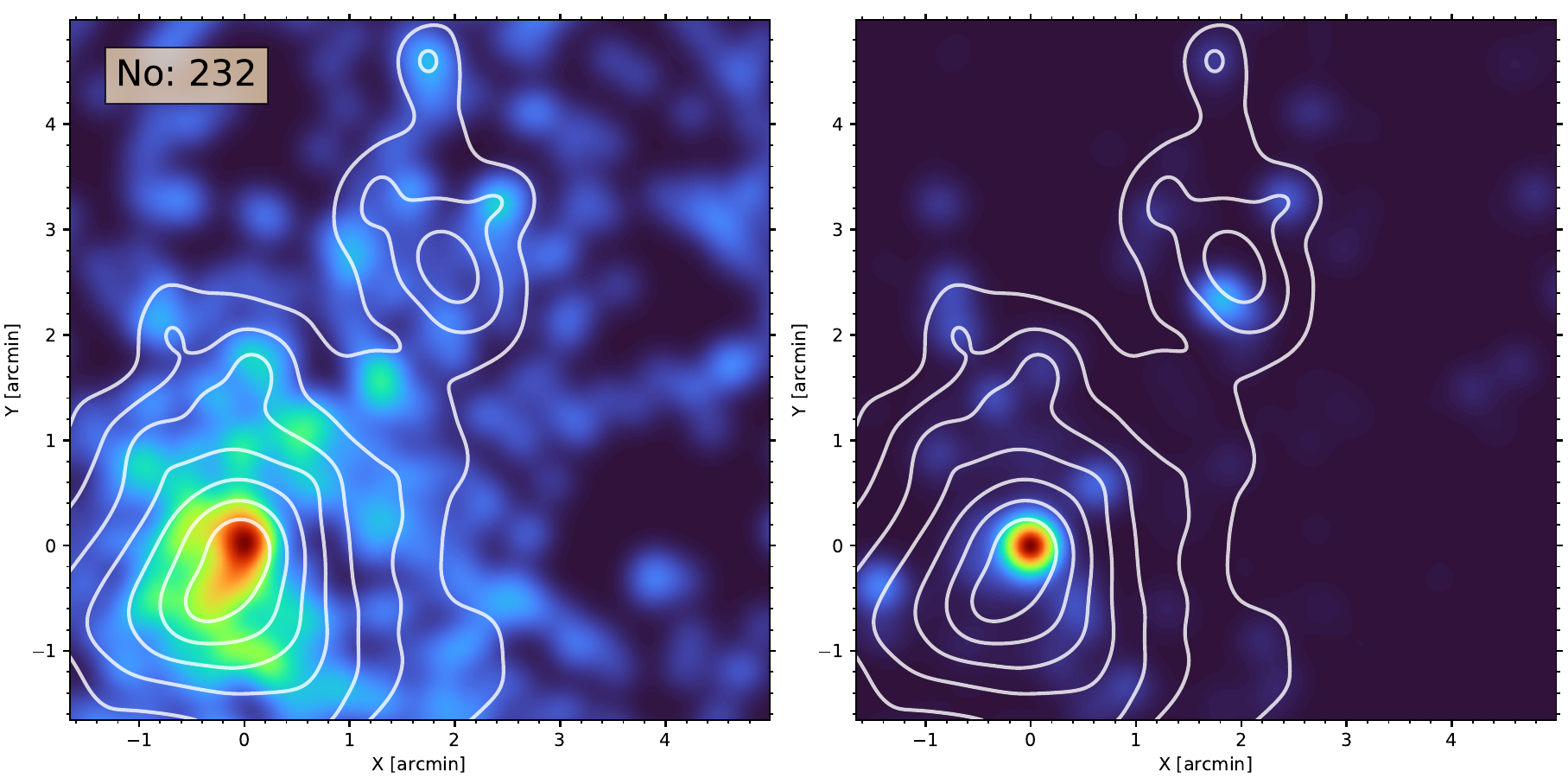}
    \includegraphics[width={0.34\textwidth}]{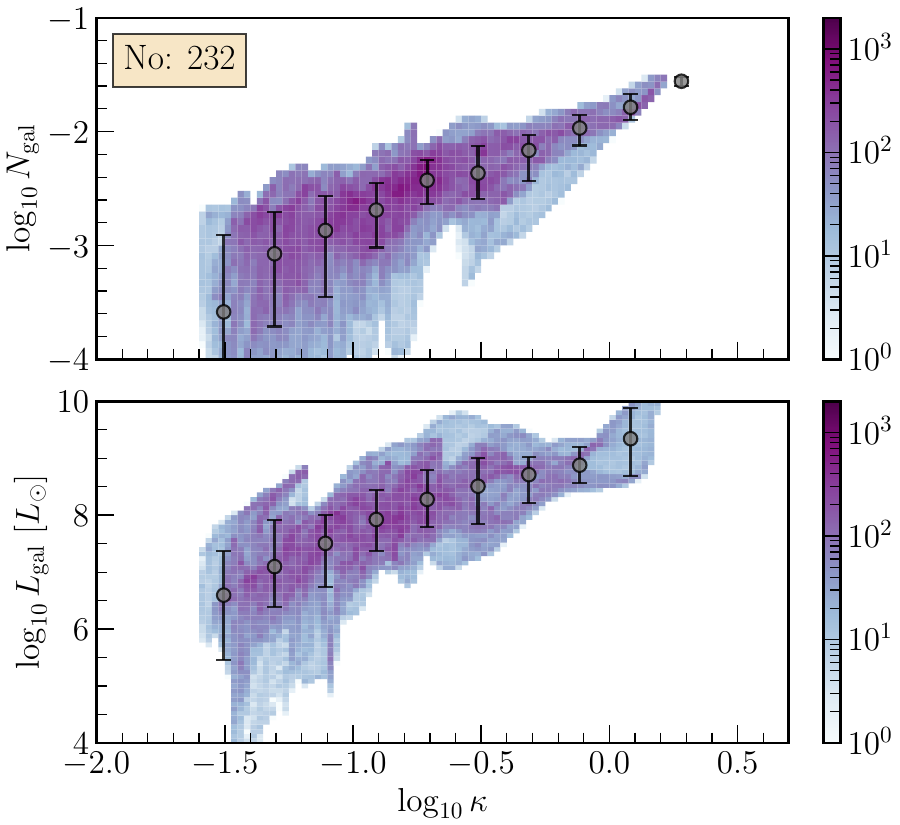}

    \includegraphics[width={0.63\textwidth}]{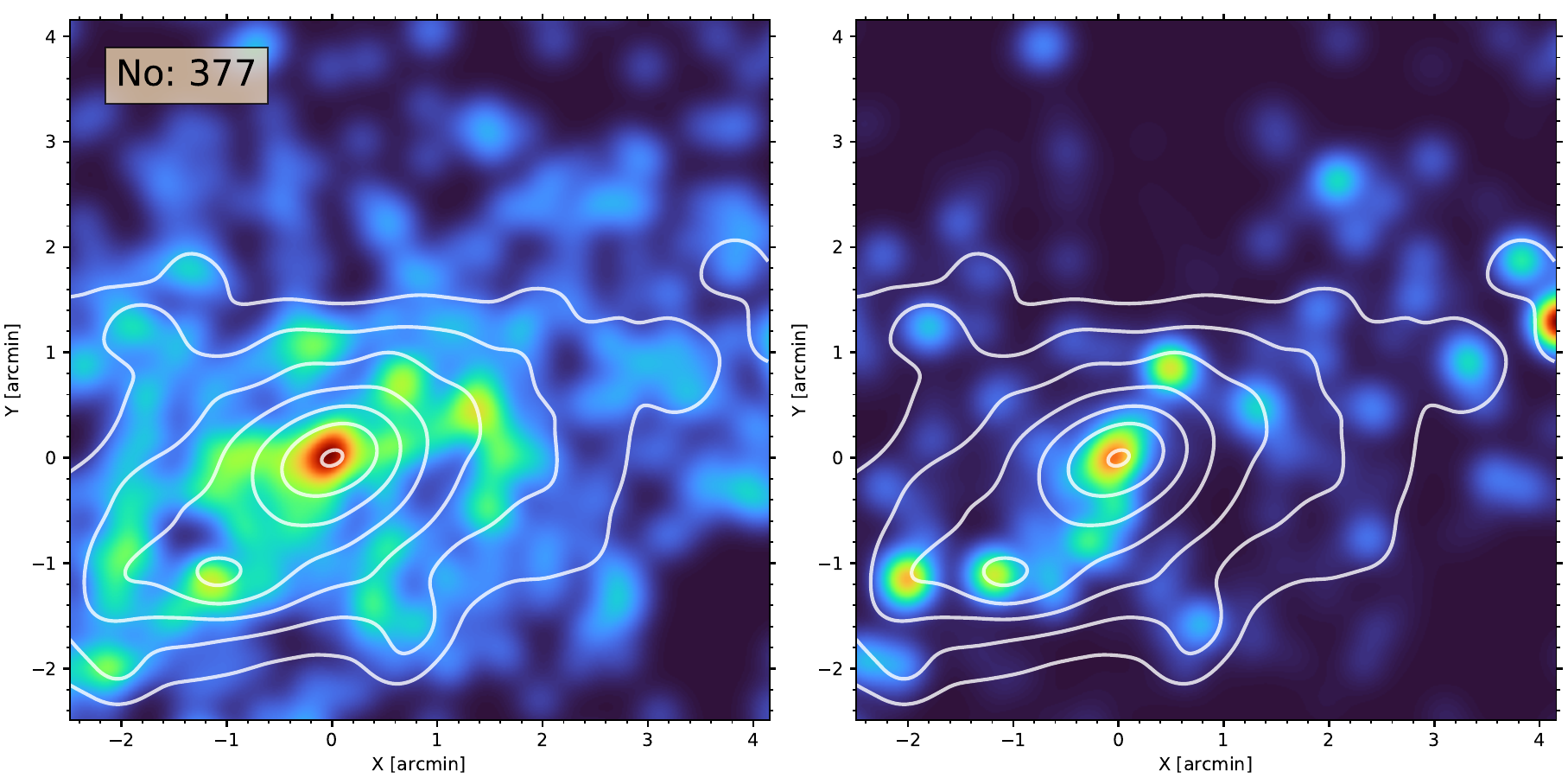}
    \includegraphics[width={0.34\textwidth}]{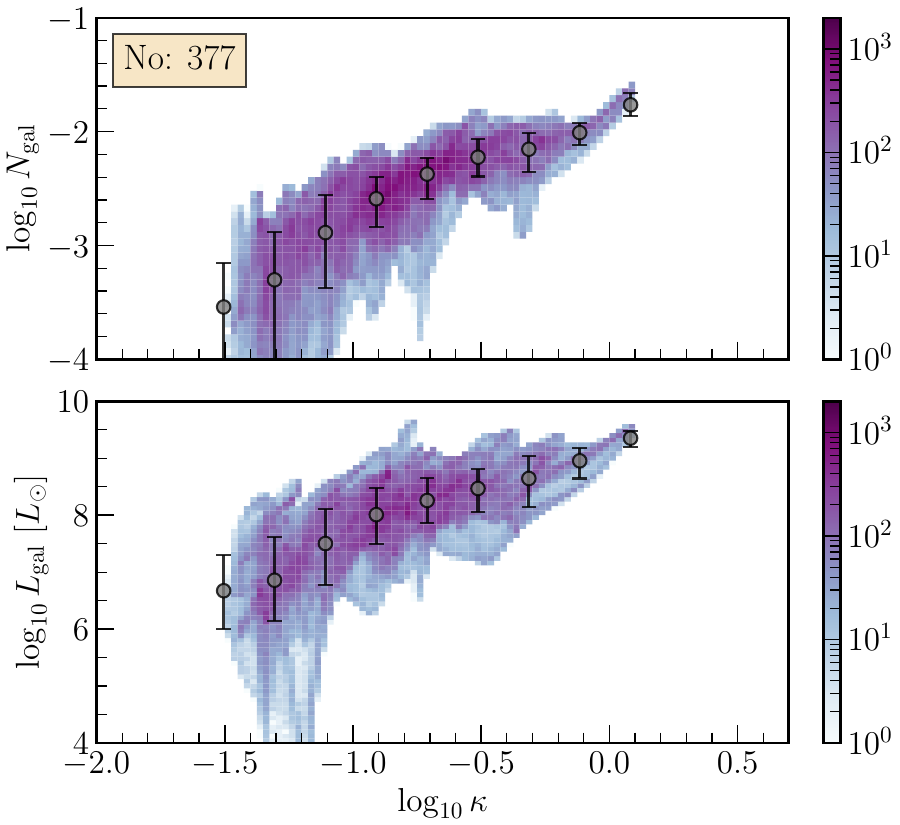}

    \includegraphics[width={0.63\textwidth}]{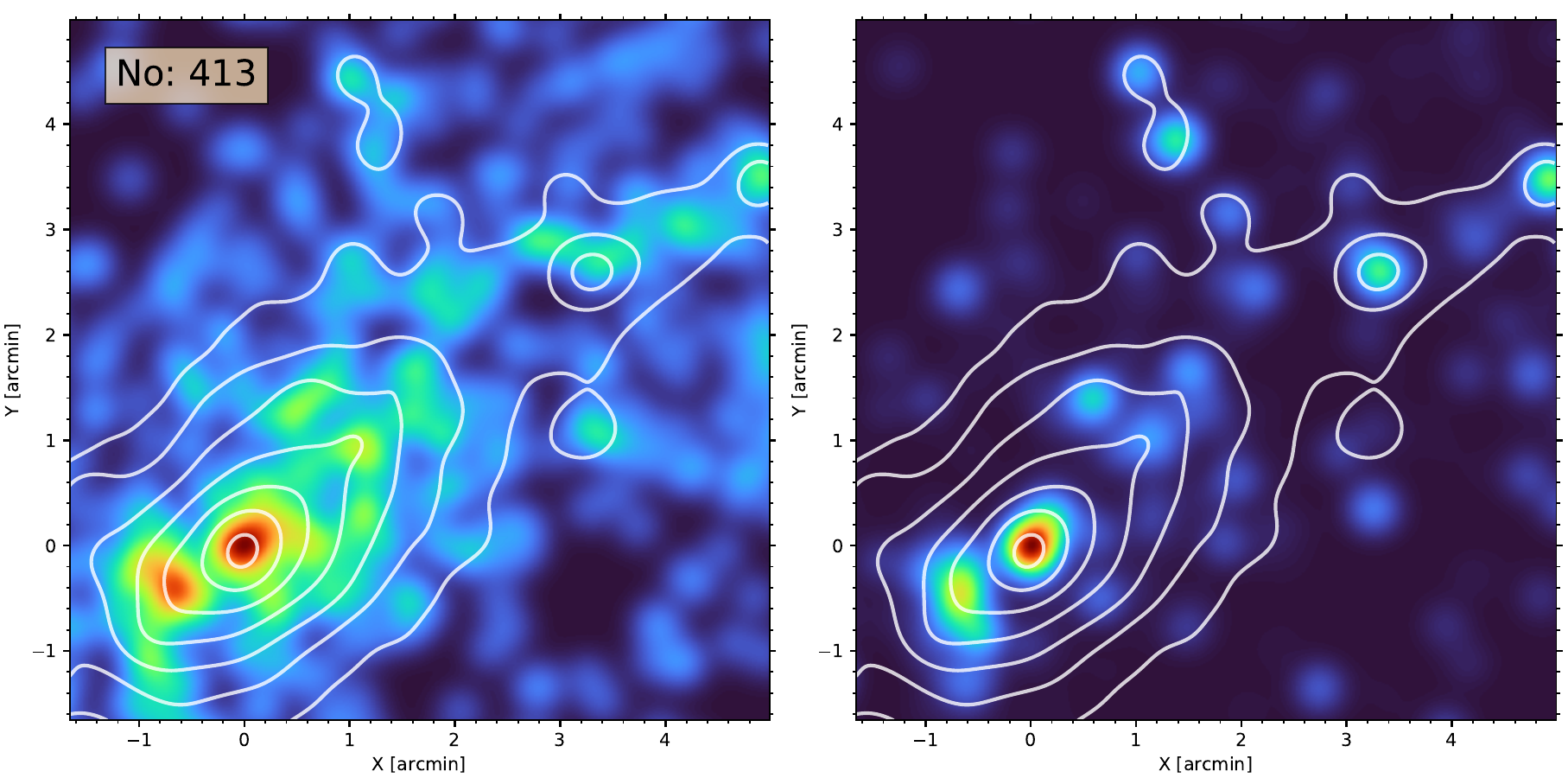}
    \includegraphics[width={0.34\textwidth}]{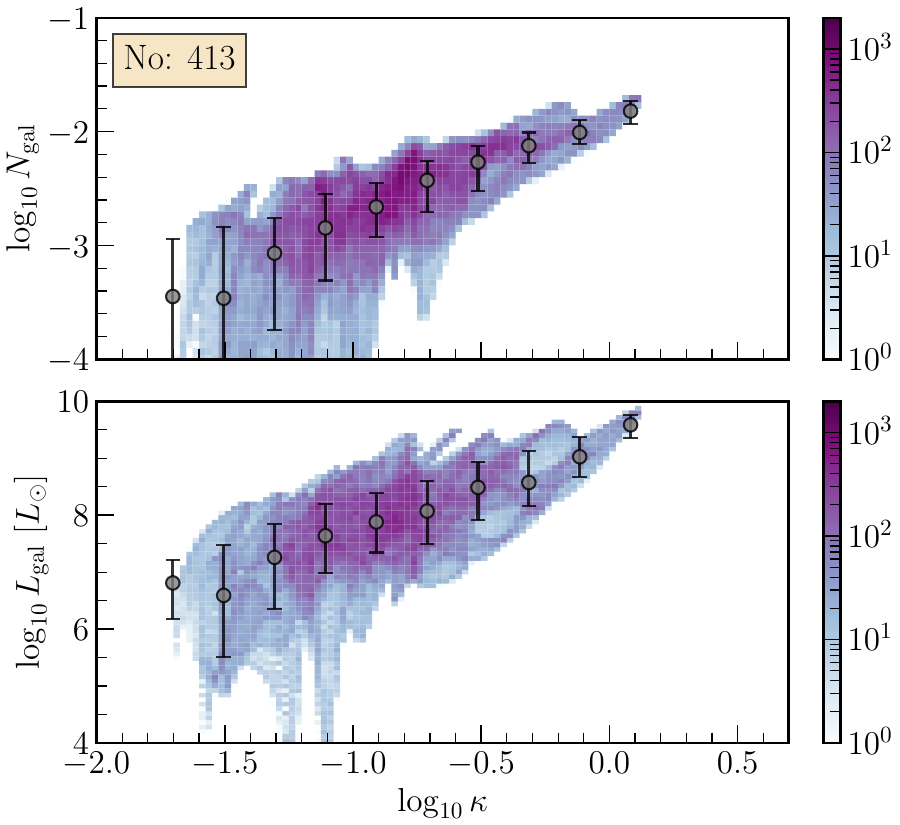}
    
    \caption{Similar to Figure~\ref{fig:HM_A2744_samplesMain}, for rest A2744 analogues in Group I. The left and middle panels show the galaxy number density map and galaxy luminosity density map, respectively. The right panel presents the two-dimensional histogram of pixel values in the cluster galaxy number/luminosity map compared to the convergence map.}
\label{fig:HM_A2744_samples_G1}.
\end{figure*}

\begin{figure*}
    \includegraphics[width={0.63\textwidth}]{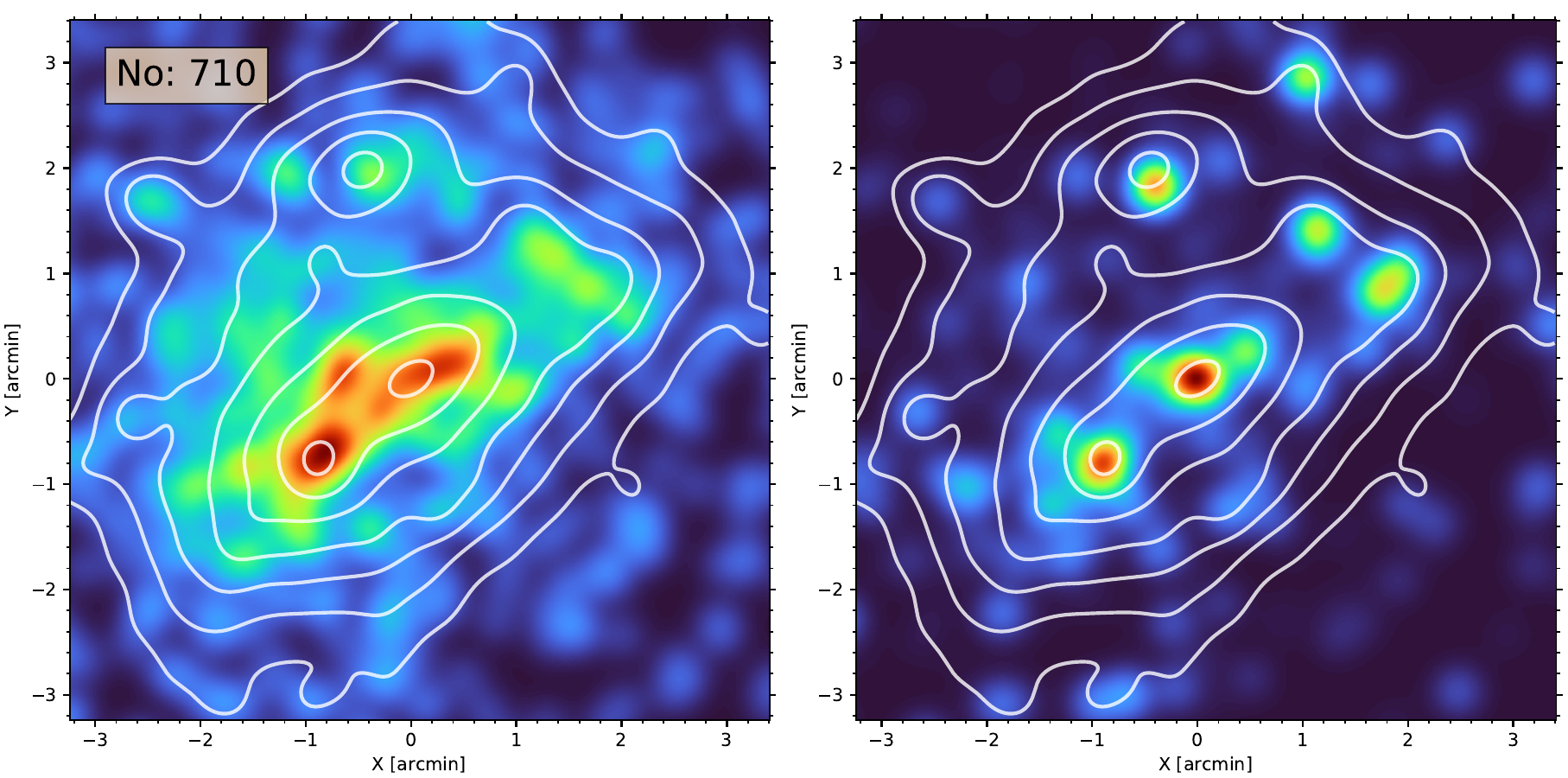}
    \includegraphics[width={0.34\textwidth}]{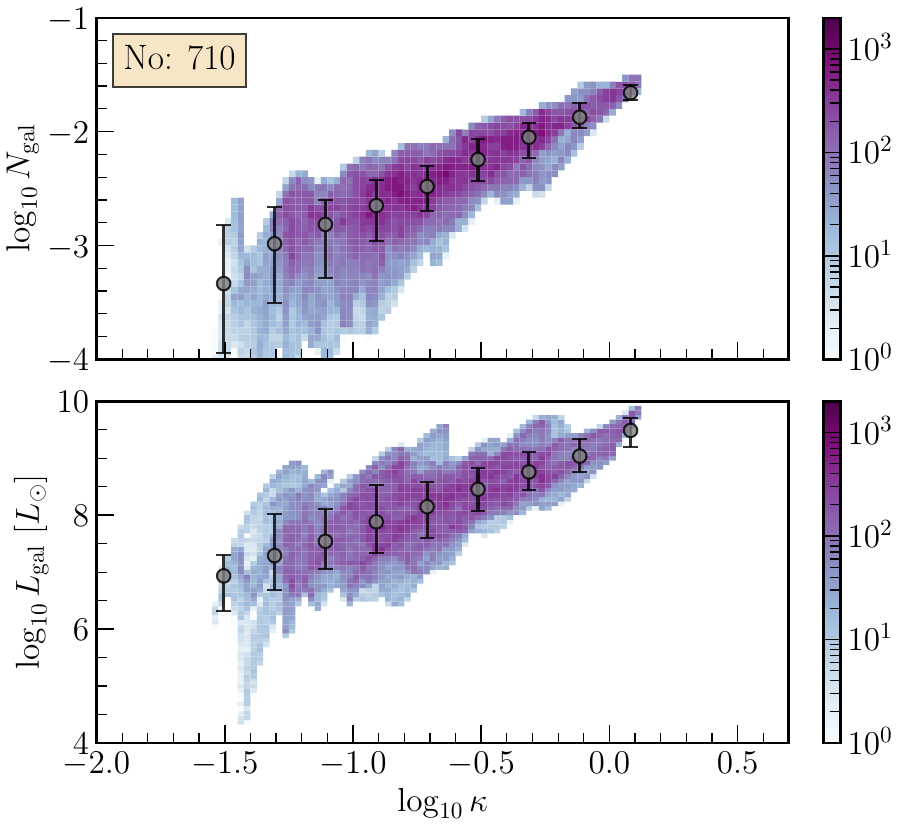}

    \includegraphics[width={0.63\textwidth}]{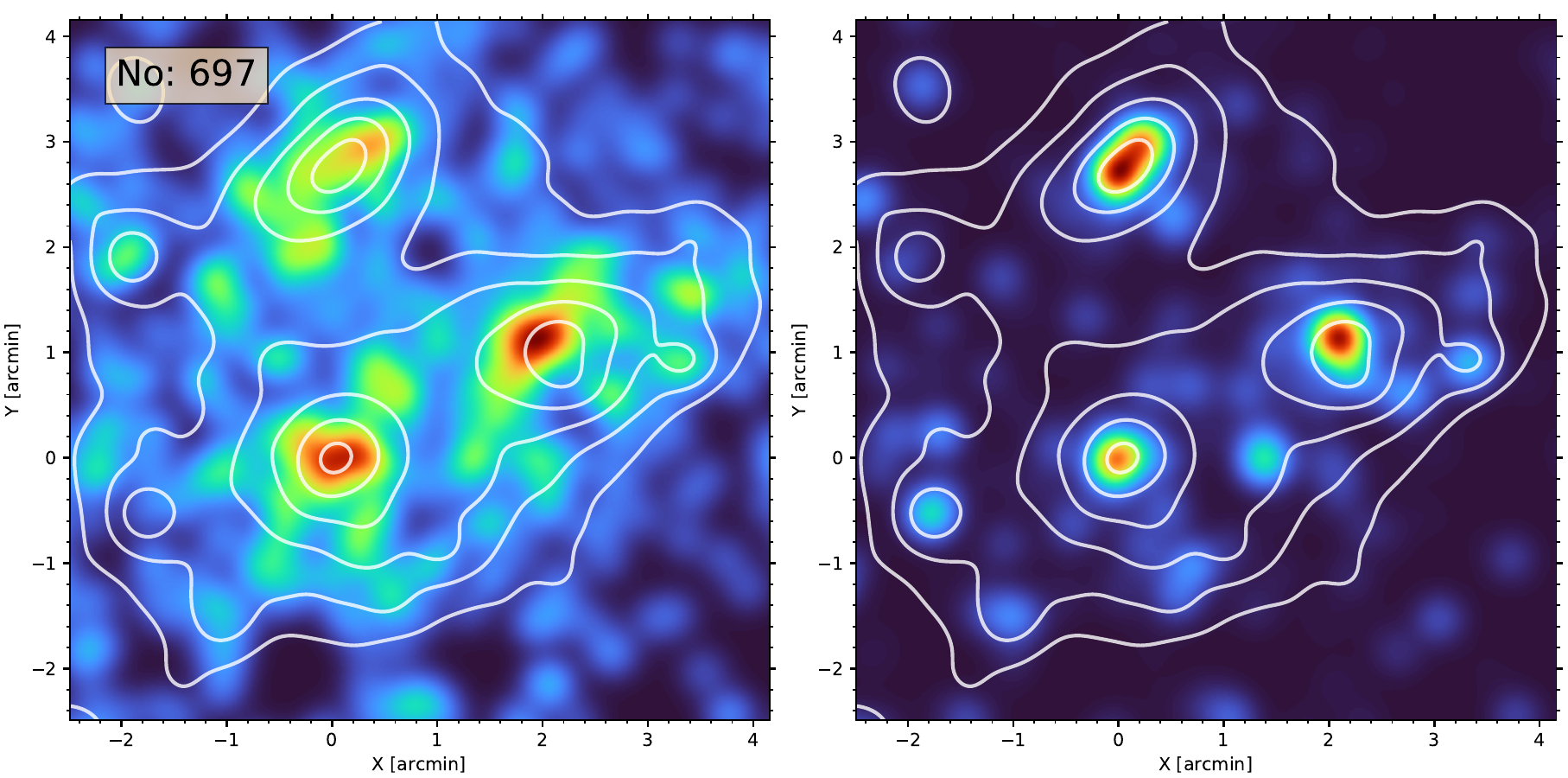}
    \includegraphics[width={0.34\textwidth}]{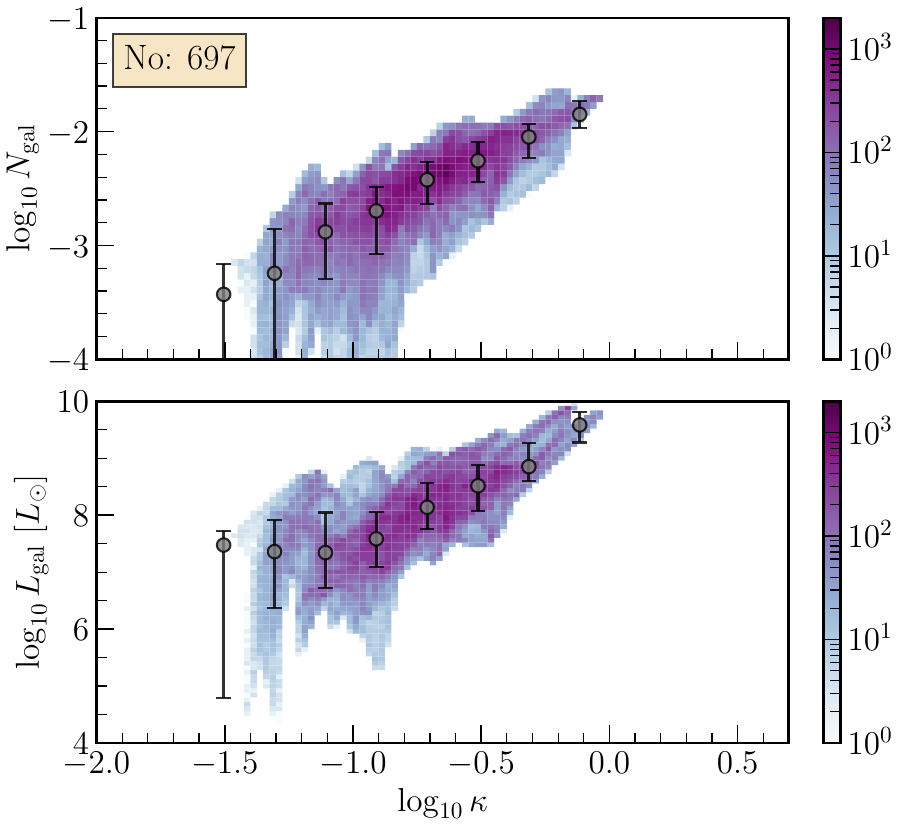}

    \includegraphics[width={0.63\textwidth}]{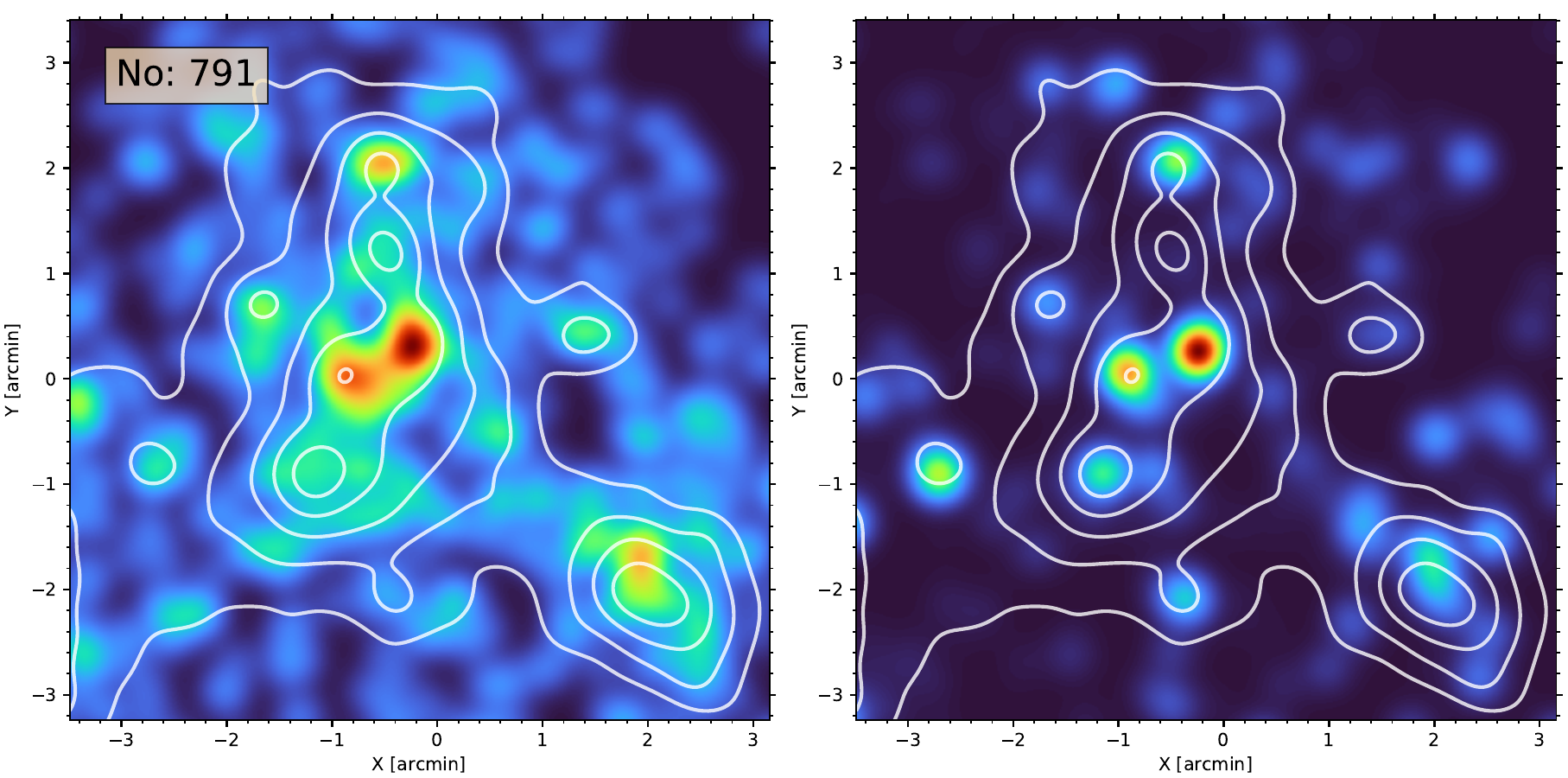}
    \includegraphics[width={0.34\textwidth}]{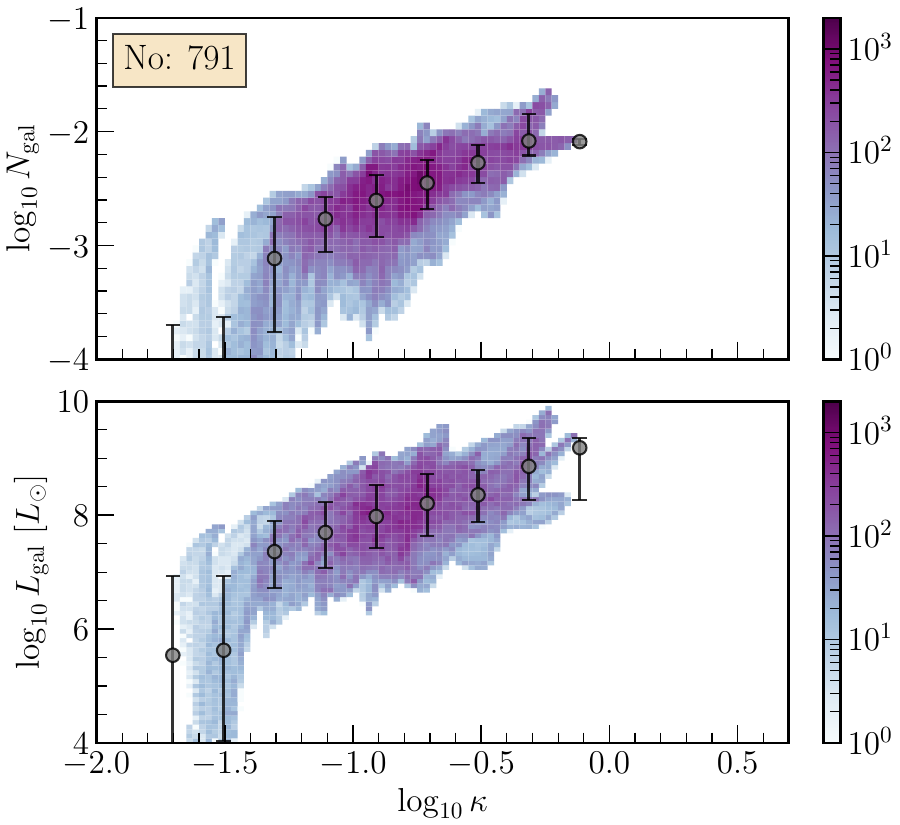}
    
    \caption{Continued to Figure~\ref{fig:HM_A2744_samples_G1}, but for rest A2744 analogues in Group II.}
\label{fig:HM_A2744_samples_G2}.
\end{figure*}


\bsp	
\label{lastpage}
\end{document}